\documentclass[superscriptaddress,aps,prl,twocolumn,showpacs,nofootinbib, longbibliography, notitlepage,floatfix]{revtex4-2}
\usepackage{etex}
\usepackage{amsmath,amssymb,amsthm}
\usepackage[colorlinks=true,citecolor=blue,urlcolor=blue]{hyperref}
\usepackage[pdftex]{graphicx}
\usepackage{times,txfonts}
\usepackage{braket}
\usepackage{comment}
\usepackage{physics}
\usepackage{color}
\usepackage{natbib}
\usepackage{amsmath,blkarray}
\usepackage{mathtools}
\usepackage{ulem}
\usepackage{latexsym}
\usepackage{tabularx, booktabs}
\usepackage{graphics,epstopdf}
\usepackage{graphicx}
\usepackage{float}
\usepackage{amsfonts}
\usepackage{tikz}
\usetikzlibrary{quantikz}
\usepackage{color,soul}

\newcommand{\be}{\begin{equation}}
\newcommand{\ee}{\end{equation}}
\newcommand{\ba}{\begin{eqnarray}}
\newcommand{\ea}{\end{eqnarray}}

\usepackage{multirow}
\usepackage{appendix}
\usepackage{url}

\begin{document}
\title{Bias-field digitized counterdiabatic quantum optimization} 

\author{Alejandro Gomez Cadavid}
\affiliation{Kipu Quantum, Greifswalderstrasse 226, 10405 Berlin, Germany}
\affiliation{Department of Physics, University of the Basque Country UPV/EHU, Barrio Sarriena, s/n, 48940 Leioa, Biscay, Spain}

\author{Archismita Dalal}
\affiliation{Kipu Quantum, Greifswalderstrasse 226, 10405 Berlin, Germany}

\author{Anton Simen}
\affiliation{Kipu Quantum, Greifswalderstrasse 226, 10405 Berlin, Germany}

\author{Enrique Solano}
\email{enr.solano@gmail.com }
\affiliation{Kipu Quantum, Greifswalderstrasse 226, 10405 Berlin, Germany}

\author{Narendra N. Hegade}
\email{narendrahegade5@gmail.com}
\affiliation{Kipu Quantum, Greifswalderstrasse 226, 10405 Berlin, Germany}

\date{\today}
\begin{abstract}
We introduce a method for solving combinatorial optimization problems on digital quantum computers, where we incorporate auxiliary counterdiabatic (CD) terms into the adiabatic Hamiltonian, while integrating bias terms derived from an iterative digitized counterdiabatic quantum algorithm. We call this protocol bias-field digitized counterdiabatic quantum optimization (BF-DCQO). Designed to effectively tackle large-scale combinatorial optimization problems, BF-DCQO demonstrates resilience against the limitations posed by the restricted coherence times of current quantum processors and shows clear enhancement even in the presence of noise. Additionally, our purely quantum approach eliminates the dependency on classical optimization required in hybrid classical-quantum schemes, thereby circumventing the trainability issues often associated with variational quantum algorithms. Through the analysis of an all-to-all connected general Ising spin-glass problem, we exhibit a polynomial scaling enhancement in ground state success probability compared to traditional DCQO and finite-time adiabatic quantum optimization methods. Furthermore, it achieves scaling improvements in ground state success probabilities, increasing by up to two orders of magnitude, and offers an average 1.3x better approximation ratio than the quantum approximate optimization algorithm for the problem sizes studied. We validate these findings through experimental implementations on both trapped-ion quantum computers and superconducting processors, tackling a maximum weighted independent set problem with 36 qubits and a spin-glass on a heavy-hex lattice with 100 qubits, respectively. These results mark a significant advancement in gate-based quantum computing, employing a fully quantum algorithmic approach.

\end{abstract}

\maketitle

{\it Introduction:}
Ising spin-glass problems are of utmost interest in both science and industry due to their vast applications. Particularly, combinatorial optimization problems, which can be formulated as solving for low-energy states of Ising spin-glass Hamiltonians, exemplify such applications \cite{lucas2014ising}. Generally, these complex optimization problems belong to the NP-hard class, making them difficult to solve on classical computers. Recent theoretical and experimental developments in this area make it a crucial topic for further exploration on current quantum computers \cite{pirnay2024principle, boulebnane2022solving, ebadi2022quantum}. A major drawback of current quantum computing hardware is its limited coherence time, connectivity, and presence of noise. These limitations pose significant challenges for widely studied quantum optimization algorithms such as the adiabatic quantum optimization (AQO) algorithm and the quantum approximate optimization algorithm (QAOA) \cite{albash2018adiabatic, farhi2014quantum}. To overcome these challenges, several alternative methods have been proposed, including the counterdiabatic (CD) protocols. In the case of AQO, CD protocols help to speed up the evolution and reduce the quantum circuit depth by suppressing non-adiabatic transitions through the addition of CD terms \cite{demirplak2003adiabatic, berry2009transitionless, del2013shortcuts}. For QAOA, CD protocols aid in designing an efficient variational circuit ansatz that quickly converges towards the solution \cite{chandarana2022digitized, chandarana2023digitized}. Despite the advantages of CD protocols, tackling large-scale problems remains a challenge, especially when it is crucial to consider higher-order CD terms, which increases the number of quantum gates.

In this work, we propose a method to tackle general Ising spin-glass instances with long-range interactions through an iterative algorithm. The output from each iteration is fed back as a bias to the input of the next iteration. This combined approach of digitized counterdiabatic quantum optimization algorithm with a bias field, called BF-DCQO, shows a drastic reduction in the time to reach both exact and approximate solutions compared to state-of-the-art approaches. This includes  DCQO~\cite{hegade2022digitized} as well as hybrid variational quantum algorithms like QAOA. Additionally, BF-DCQO does not require any classical optimization subroutines, thus overcoming the trainability issues faced by variational quantum optimization algorithms \cite{wang2021noise}. We experimentally demonstrate the potential of the proposed method on trapped-ion quantum computers with up to fully-connected 36 qubits and superconducting quantum processors with sparsely connected 100 qubits. We will discuss the advantages of the proposed BF-DCQO method in the light of competing protocols for approaching quantum advantage for industry use cases.
\begin{figure*}
    \centering
\includegraphics[width=0.9\linewidth]{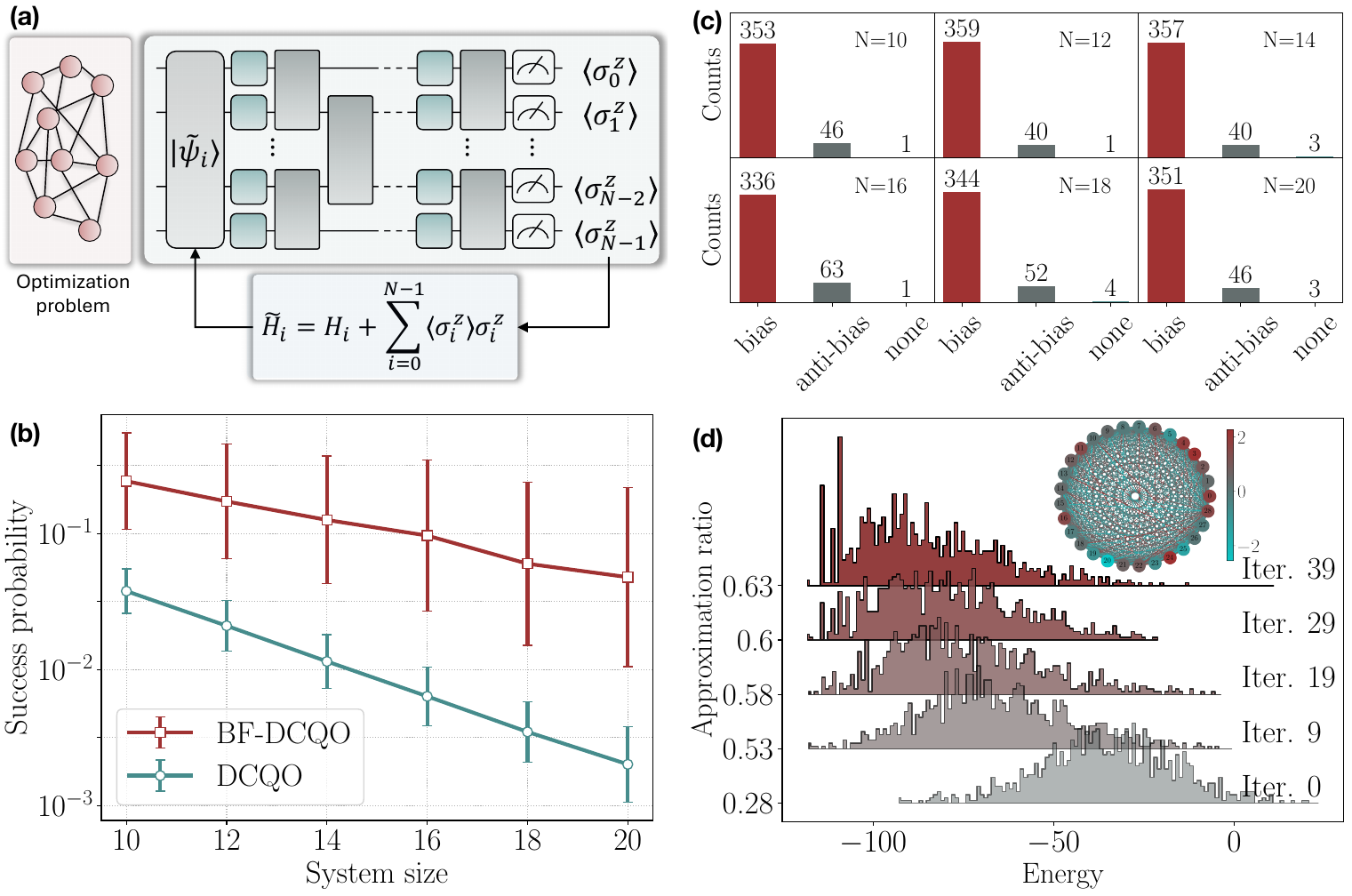}
    \caption{BF-DCQO for the Ising spin-glass problem with all-to-all interaction. In (a), the schematic of the BF-DCQO procedure is shown. In (b), the ground state success probability is plotted for system sizes ranging from 10 to 20 qubits. For each system size, 400 randomly generated spin-glass instances are taken from a normal distribution with a mean of $0$ and a variance of $1$. We present the scaling of the BF-DCQO algorithm with 10 iterations and the standard DCQO, with simulation parameters $\Delta t = 0.1$ and $n_{\text{trot}} = 3$. The error bars represent the standard deviation of the random instances. In (c), the classification of the 400 instances using enhancement in success probability as a criterion, counting the number of instances successfully tackled by either the bias or anti-bias field DCQO, and those that failed in both cases. In (d), emulation results for a randomly generated spin-glass problem with 29 qubits are shown. We performed 39 iterations of BF-DCQO, displaying the increasing approximation ratio on the y-axis. For each iteration, we used $n_{\text{trot}} = 2$, $n_{\text{shots}} = 1000$, and the {\it IonQ-Forte 1} noise model, accessed through IonQ-cloud \cite{IonQCloud2023}. Additionally, the associated all-to-all connected graph is depicted.}
    \label{fig1}
\end{figure*}

{\it DCQO Algorithm--}
An adiabatic quantum optimization protocol to find the ground state of an Ising spin-glass problem with all-to-all connectivity is described by the Hamiltonian \(H_{\text{ad}}(\lambda)=  [1-\lambda(t)] H_i + \lambda(t) H_f\). Here, \(H_i\) is the initial Hamiltonian whose ground state can be easily prepared, typically chosen as a transverse field \(H_i = - \sum_{i=1}^{N} \sigma_{i}^{x}\) with the ground state \(|+\rangle^{\otimes N} \), where $N$ is the number of spins. The final Hamiltonian, corresponding to the spin-glass, is \(H_f  = \sum_{i<j}^{N} J_{ij} \sigma_{i}^{z} \sigma_{j}^{z}+\sum_{i=1}^{N} h_{i} \sigma_{i}^{z}\).
Here, \(\lambda\) is a time-dependent control function that describes the adiabatic path. In the adiabatic limit, i.e., \(\dot{\lambda}(t) \to 0\), the system follows the instantaneous eigenstates. However, in practice, following the slow adiabatic evolution is affected by hardware noise and the limited coherence time. On the other hand, fast evolution results in non-adiabatic transitions. To overcome this challenge, CD protocols have been proposed \cite{demirplak2003adiabatic, berry2009transitionless}. In CD protocols, the idea is to introduce an auxiliary velocity-dependent (\(\dot{\lambda}(t)\)) term to the Hamiltonian to suppress non-adiabatic transitions. This takes the form
\begin{equation}
   H_{\text{cd}}(\lambda)=  H_{\text{ad}}(\lambda)  + \dot{\lambda} A_{\lambda}, 
\end{equation}
where \(A_{\lambda}\) is known as the adiabatic gauge potential \cite{kolodrubetz2017geometry}. Obtaining and realizing \(A_{\lambda}\) is a highly resource-demanding task for many-body Hamiltonians. Rather, there are several proposals to obtain this in an approximate way \cite{sels2017minimizing, claeys2019floquet, hatomura2021controlling,xie2022variational, takahashi2024shortcuts}. We consider the nested commutator method where the adiabatic gauge potential can be written as the series expansion \(A_{\lambda}^{(l)} = i\sum_{k=1}^l \alpha_k(t) \mathcal{O}_{2k-1} (t)\). Here, \(l\) is the expansion order, and the operator \(\mathcal{O}_{k}(t) = [ H_{\text{ad}}, \mathcal{O}_{k-1}(t) ]\) with \(\mathcal{O}_{0}(t) = \partial_\lambda H_{\text{ad}}\). In the limit \(l \to \infty\), the exact \(A_{\lambda}\) can be obtained. The CD coefficient \(\alpha_k(t)\) can be calculated by variational minimization as detailed in the supplementary material \cite{supplement2024}. To solve for the Ising spin-glass problem, we simply consider the first-order approximation as \(A_\lambda^{(1)} = - 2 \alpha_1 \left[ \sum_{i=1}^{N} h_i \sigma^y_i + \sum_{i<j}^{N} J_{i j}\left(\sigma^y_i \sigma^z_j+\sigma^z_i \sigma^y_j\right) \right]\). The time evolution of the Hamiltonian in the given equation, even with the first-order CD term, is a challenging task on current analog quantum processors due to a lack of flexibility. Also, an important fact to notice is that the obtained CD terms are non-stoquastic with off-diagonal matrix entries being imaginary. Realization of such terms on current quantum annealers is unfeasible. To overcome this challenge, digitized counterdiabatic quantum protocols have been proposed to realize the CD protocols on gate-model quantum computers \cite{hegade2021shortcuts}. Not only does the digital approach provide the flexibility to realize arbitrary CD terms, it also helps to further improve the CD protocols because of the flexibility in the control parameters. To realize the time evolution of the CD Hamiltonian, we use the first-order product formula \cite{barends2016digitized} with a number of Trotter steps \(n_{\text{trot}}\), step size \(\Delta t\), and total evolution time \(T\). The unitary describing the evolution is given by
\begin{equation}
|\psi(T)\rangle=\left[\prod_{k=1}^{n_{\text{trot}}} \prod_{j=1}^{n_\text{terms}} \exp \left\{-i \Delta t \gamma_j(k \Delta t) H_j\right\} \right]  |\psi_i\rangle.
\end{equation}
Here, \(|\psi_i\rangle\) is the initial ground state and \(H_{\text{cd}} = \sum_{j=1}^{n_\text{terms}} \gamma_j(t) H_j\), where \(n_\text{terms}\) is the number of local Pauli operators \(H_j\). Each product of matrix exponentials is decomposed into quantum gates with one and two-qubit gates. Even with the first-order CD approximation, a polynomial scaling enhancement in ground state success probability has been shown in comparisons to finite time AQO~\cite{hegade2022digitized, hartmann2022polynomial}. Going for higher-order CD terms can improve the results further but this comes at the cost of additional quantum gates. The main advantage of this approximate CD protocol is that, even with a very short evolution time or circuit depth, one can obtain low energy states in comparison to digitized adiabatic evolution. To obtain similar performance, one would require very long-depth digitized adiabatic evolution, which is not feasible because of the limited coherence time and the noise. This main feature is the key to developing the concept of bias-field for the CD protocol.
\begin{figure}
    \centering
    \includegraphics[width=1\linewidth]{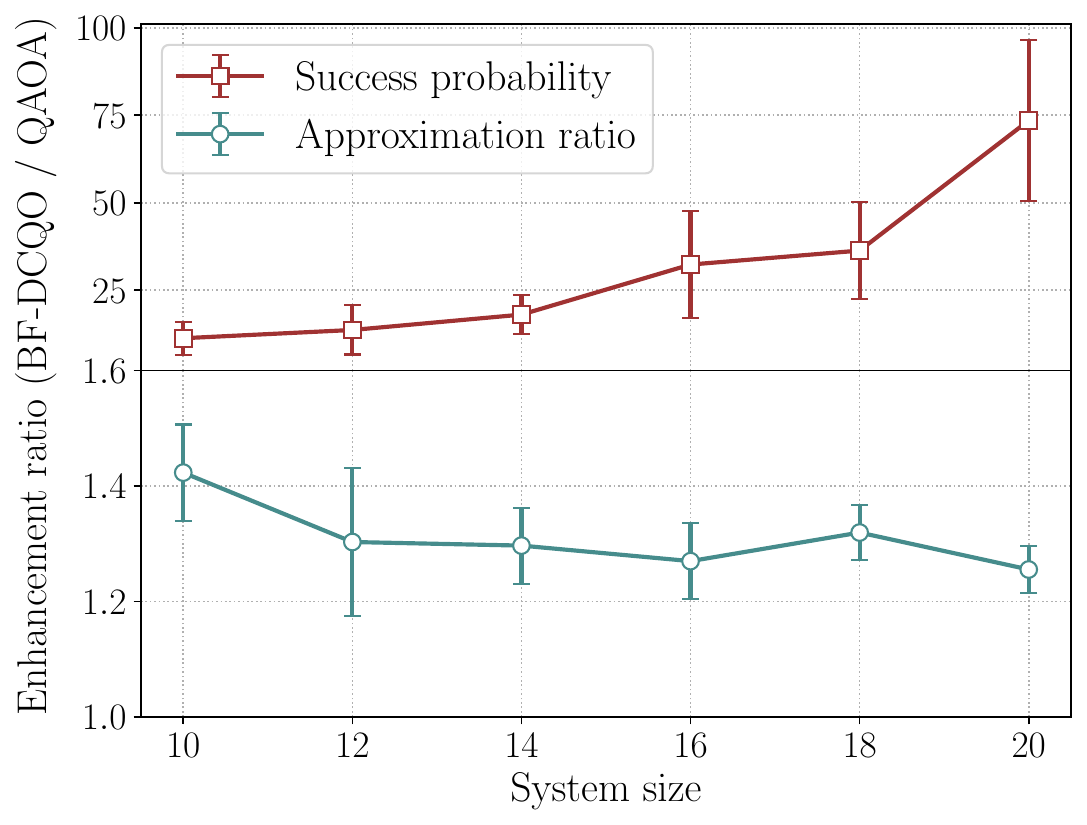}
    \caption{Comparison between simulated BF-DCQO and QAOA ($p=3$). The analysis was conducted on 10 different random all-to-all connected spin-glass instances, with system sizes ranging from 10 to 20 qubits. For QAOA, we used 20 different random initializations with the COBYLA optimizer, setting the maximum number of iterations at 300. This was contrasted with 10 iterations of BF-DCQO, where $n_{\text{trot}} = 3$.}
    \label{fig2}
\end{figure}
\begin{figure*}
    \centering
  \includegraphics[width=0.9\linewidth]{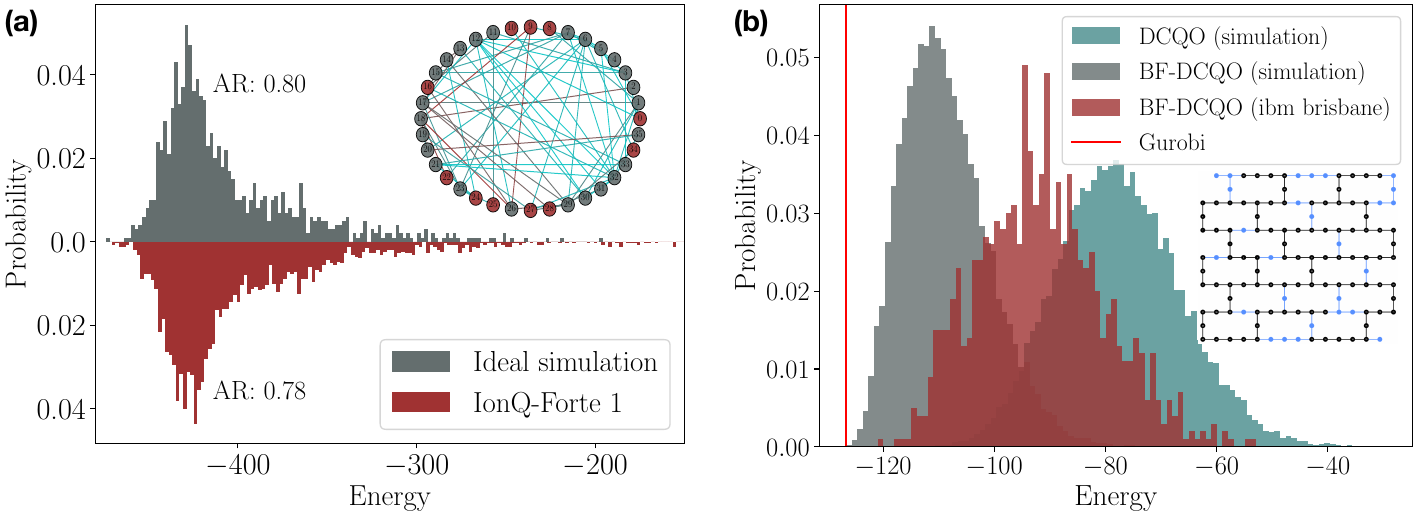}
    \caption{Experimental results: In (a), the ninth iteration of BF-DCQO for a randomly generated 36-node weighted MIS instance is shown. For the simulations, we used $n_{\text{trot}} = 3$, and $n_{\text{shots}} = 1000$. In the experimental case on {\it IonQ-Forte 1}, only the ninth iteration was run, using $n_{\text{shots}} = 2500$ with error mitigation. The MIS size was 16, and we obtained an independent set of size 11, as depicted in the graph. In (b), the eighth iteration of BF-DCQO for a nearest-neighbor randomly generated 100-qubit spin-glass instance is displayed. For these simulations, we used $n_{\text{trot}} = 2$ and $n_{\text{shots}} = 25000$. For the experimental results on {\it ibm\_brisbane}, $n_{\text{shots}} = 1000$ was used. Additionally, the circuit layout on the hardware is shown.}
    \label{fig3}
\end{figure*}

{\it DCQO with longitudinal bias-field--} Initialization plays a crucial role in the success of AQO and DCQO. Instead of starting with a random initial state, beginning with a good one that encodes some information about the final solution is beneficial \cite{grass2019quantum, grass2022quantum}. Several proposals exist for such warm starting techniques, wherein inexpensive classical methods are used to solve a relaxed version of the problem and then utilize this solution as an input for the quantum algorithm~\cite{egger2021warm, truger2023warm}. These techniques inherit the performance guarantee of the classical algorithm.
Then, we propose a different approach where the solution from DCQO is fed back as a bias to the input state for the next iteration. The total Hamiltonian, which includes the longitudinal bias field, is defined as
\begin{align} \label{H_total}
     H(\lambda)&=  [1-\lambda(t)] \tilde {H_i} + \lambda(t) H_f + \dot{\lambda} A_{\lambda}^{(l)},\\
&\text{with} \; \tilde{H_i} = \sum_{i=1}^{N} \left [h_i^x \sigma^x_i - h_i^b \sigma^z_i \right ]  \nonumber ,
\end{align}
where the value of the longitudinal bias field, \( h_i^b = \langle \sigma^z_i \rangle \), is obtained by measuring the qubits in the computational basis in each iteration. Since the solution from the first iteration of DCQO is expected to sample low-energy states of the spin-glass Hamiltonian, the Pauli-$Z$ expectation value can serve as an effective bias for the next iteration, steering the dynamics toward the actual solution. Because the bias term alters the initial Hamiltonian and, consequently, the initial ground state, the new ground state \( |\tilde{\psi_i}\rangle \) must be used as the input for the next iteration. The smallest eigenvalue of the single-body operator \( \left[h_i^x \sigma^x_i - h_i^b \sigma^z_i \right ]  \) is given by \( \lambda^{\min}_i = -\sqrt{(h^b_i)^2 + (h^x_i)^2} \), and its associated eigenvector is \( \ket{\tilde{\phi}}_i = \phi_0 \begin{bmatrix} 1 & \frac{h^b_i + \lambda^{\min}_i}{h^x_i} \end{bmatrix}^\intercal \). Since \( h^b_i \) and \( h^x_i \) are projections on the z- and x-axes, the corresponding ground state can be prepared by a y-axis rotation given by \( \theta_i = \tan^{-1}\left(\frac{h^x_i}{h^b_i + \sqrt{(h^b_i)^2 + (h^x_i)^2}}\right) \). Therefore, the ground state of \( \tilde{H_i} \) can be prepared using $N$ y-axis rotations as \( |\tilde{\psi_i}\rangle = \bigotimes_{i=1}^{N} |\tilde{\phi} \rangle_i = \bigotimes_{i=1}^{N} R_y(\theta_i)|0\rangle_i \). In Fig~\ref{fig1}(a), the schematic diagram depicting the BF-DCQO is shown.

To analyze the performance of the BF-DCQO, we consider 400 random instances of the spin-glass problem, with coupling \( J_{ij} \) and \( h_i \) obtained from a Gaussian distribution with a mean of 0 and a variance of 1. The scheduling function is \( \lambda(t) = \sin^2\left[\frac{\pi}{2}\sin^2\left(\frac{\pi t}{2 T}\right)\right] \) and we use \( h^x_i = -1 \). We only consider first-order CD terms and the CD coefficient changes during each iteration since the initial Hamiltonian changes. In this case, we analytically calculate the exact form of the CD coefficient \( \alpha_1 \) for the Hamiltonian in Eq.~\eqref{H_total}. Fig~\ref{fig1}(b) illustrates the ground state success probability with increasing system size for BF-DCQO with 10 iterations and naive DCQO, given a fixed evolution time \( T \) with three Trotter steps. In both cases, the success probability $p_{gs}=\left|\left\langle\psi_{gs} \mid \psi_f(T)\right\rangle\right|^2$, where $\ket{\psi_{gs}}$ is the actual ground state of the spin-glass Hamiltonian, decreases exponentially with system size. However, the exponential factor for BF-DCQO is smaller than that for DCQO, indicating a polynomial scaling advantage. We noticed that not all 400 instances are improved by the inclusion of the bias field. In some cases, if the solution from the first iteration of DCQO leads to undesired outcomes, employing an anti-bias field \( h_i^b = -\langle \sigma^z_i \rangle \) may help suppress these outcomes. In Fig~\ref{fig1}(c), the number of instances enhanced by the bias field is depicted. For the unsuccessful instances, employing an anti-bias field shows improvement. However, there are few instances where both the bias and anti-bias fields fail. This is primarily due to the simulation parameters we have chosen in this work, and altering them may lead to successful results. 

To evaluate the performance of the algorithm in the presence of hardware noise, we use a noisy emulator that mimics the actual noise model of a trapped-ion hardware system, {\it IonQ Forte}. We consider a fully connected 29-qubit spin-glass instance and implement the BF-DCQO algorithm. The energy distribution across each iteration is shown in Fig~\ref{fig1}(d). Remarkably, even with just \( n_{\text{trot}}=2 \) steps and the number of shots \( n_{\text{shots}} = 1000 \), the algorithm guides the dynamics toward the solution. By iteration 29, the exact ground state was obtained. Additionally, it is clear that the approximation ratio \( E_{\text{obtained}} / E_{\text{gs}} \) improves with each iteration.

An important aspect of BF-DCQO is that it does not require any classical optimization subroutines as in variational quantum algorithms (VQA). This feature makes it an impressive approach, as the main drawback of VQA lies in trainability issues like barren plateaus and local minima. The presence of noise makes it even harder to rely on VQAs. Since we have already seen the successful performance of BF-DCQO in noisy conditions, we compare its performance with QAOA, a widely used variational quantum optimization algorithm. We consider 10 random instances of the long-range spin-glass problem across various sizes. Ground-state success probability and approximation ratio are used as metrics for comparison.

To maintain the same circuit depth, we consider QAOA with $p=3$ layers and BF-DCQO with \( n_{\text{trot}} = 3 \). For optimizing the QAOA circuit, we use the COBYLA optimizer with a maximum of 300 iterations. For each instance, the best solution out of 20 random initializations is considered for QAOA.  For BF-DCQO, we employ just 10 iterations of feedback. In Fig~\ref{fig2}, we plot the success probability enhancement ratio, which is the ratio of the ground state success probability obtained with BF-DCQO versus QAOA, as well as the enhancement ratio of the approximation ratio. Despite requiring two orders of magnitude fewer iterations, BF-DCQO outperforms QAOA in both metrics. Moreover, the success probability enhancement ratio increases with system size, showing a 75x improvement for the 20-qubit case. On average, we observe a 1.3x improvement in the approximation ratio with BF-DCQO.

{\it Experimental implementation:}
For the experimental validation of BF-DCQO, we consider a 36-qubit trapped-ion quantum processor, {\it IonQ Forte}, and a 127-qubit superconducting quantum processor, {\it ibm\_brisbane}. We explore two problems that can be suitably mapped to the hardware connectivity: a randomly generated Weighted Maximum Independent Set (WMIS) problem with 36 nodes, implemented on trapped-ion hardware, and an instance of the Ising spin-glass problem on a heavy-hex lattice with 100 spins, implemented on superconducting hardware.

The WMIS is a combinatorial optimization problem where the objective is to identify a subset of vertices in a graph that are mutually non-adjacent (an independent set) and have the highest possible total weight. Given a graph \( G = (V, E) \) with vertices \( V \) and edges \( E \), each vertex \( v \in V \) has an associated weight \( w(v) \). The task is to find a subset of vertices \( I \subseteq V \) such that no two vertices in \( I \) are connected by an edge in \( E \), while maximizing the sum of the weights of the selected vertices,
\begin{align}
    \text{Maximize}& \quad \sum_{v \in I} w(v), \notag\\
    \text{subject to} \quad (u, v)& \notin E \quad \forall \, u, v \in I. \notag
\end{align}
This problem is NP-hard because it generalizes the classic MIS problem by incorporating vertex weights. The WMIS problem can be mapped to the Ising spin-glass Hamiltonian by associating a binary spin variable with each vertex, and defining interactions that penalize adjacent vertices that are both included, while rewarding vertices that are selected based on their weights.
Since in the WMIS problem the interactions between the qubits can be long-range, trapped-ion systems are well suited to tackle this problem without requiring any SWAP gates. In Fig~\ref{fig3}(a), the experimental result from {\it IonQ Forte} for a 36-node WMIS is shown. To optimize access to the hardware, we first ran the BF-DCQO on an ideal local simulator and then ran only the final circuit corresponding to the last iteration on the hardware. The error-mitigated experimental result is in close agreement with the ideal simulation result. In the experiment, we considered \( n_{\text{trot}} = 3 \), \( n_{\text{shots}} = 2500 \), and used hardware native gates for circuit implementation. Additionally, we performed debias error mitigation \cite{maksymov2023enhancing} and circuit optimization to reduce the total gate counts further. For the considered WMIS problem, the maximum independent set size is 16, and the obtained independent set size from the experiment is 11.

As a second example, we consider a spin-glass problem on a heavy hexagonal lattice. Since the interaction terms in the problem Hamiltonian match the hardware connectivity, we can consider a large system size of 100 qubits on {\it ibm\_brisbane} hardware. In Fig~\ref{fig3}(b), we show the ideal simulation results for DCQO and BF-DCQO, and the experimental result for BF-DCQO. We also consider a classical solver, Gurobi~\cite{gurobi}, as a reference. We notice that, even with just 10 iterations, BF-DCQO provides a drastic enhancement compared to DCQO. Additionally, in the absence of noise, BF-DCQO reaches the solution obtained from Gurobi with just two Trotter steps. Although the experimental results are slightly different from the ideal result due to noise, the performance is better than the ideal DCQO. More details on the experimental implementation can be found in the supplementary information~\cite{supplement2024}. 

{\it Discussion and conclusion:} 
We introduced BF-DCQO, an iterative quantum optimization algorithm designed to tackle combinatorial optimization problems mapped to long-range Ising spin-glass problems. By feeding back the solution from each iteration as the input for the next one, BF-DCQO incrementally refines the initial ground state, bringing it progressively closer to the final ground state. This iterative approach, combined with CD protocols that prepare low-energy states using short-depth quantum circuits, makes BF-DCQO well-suited for large-scale combinatorial optimization problems on current quantum hardware with limited coherence times.

Our simulation results demonstrate a polynomial scaling advantage in ground-state success probability compared to finite-time digitized AQO and DCQO for fully connected spin-glass problems. Additionally, noisy simulations with realistic noise models for a fully connected 29-qubit spin-glass problem showcase the algorithm's robustness, achieving exact ground states despite the presence of noise. The absence of classical optimization subroutines in BF-DCQO helps mitigate trainability issues commonly associated with VQAs. Comparisons with the QAOA reveal significant enhancements in ground-state success probability and approximation ratios while requiring fewer computational resources. Although BF-DCQO shows great promise, as a purely quantum algorithm, future work could also explore hybrid versions incorporating variational parameters and higher-order CD terms. Experimental validation on a 36-qubit trapped-ion quantum computer and a 100-qubit superconducting quantum computer confirmed good agreement with ideal simulations. Looking ahead, BF-DCQO could tackle more challenging instances of long-range spin-glass problems on next-generation trapped-ion hardware with over 60 qubits, potentially providing empirical evidence of quantum speed-up by comparing its performance with classical algorithms.  
\bibliography{reference.bib}

\pagebreak
\onecolumngrid

\vspace{1cm}

\begin{center}
\textbf{\large Supplemental Materials: Bias-field digitized counterdiabatic quantum optimization}
\end{center}
\setcounter{equation}{0}
\setcounter{figure}{0}
\setcounter{table}{0}
\setcounter{page}{1}
\makeatletter
\renewcommand{\theequation}{S\arabic{equation}}
\renewcommand{\thefigure}{S\arabic{figure}}
\renewcommand{\bibnumfmt}[1]{[#1]}
\renewcommand{\citenumfont}[1]{#1}
%

In this supplementary information, we provide the analytical calculation of the CD coefficient, details about the simulation and experimental procedures, and additional results to support the concepts discussed in the main article.

\section{Analytical calculation of the first order counterdiabatic term}
 
The $\ell$-th order AGP approximation of a system described by the adiabatic Hamiltonian $H_{\text{ad}}(\lambda)$, is given by \cite{hatomura2024shortcuts}
\begin{equation}
\label{eq:AGP}
    A_\lambda^{(\ell)} = i \sum_{k=1}^{\ell} \alpha_k(\lambda) O_{2k-1}(\lambda),
\end{equation}
where $O_{k+1}(\lambda) = \comm{H_{\text{ad}}(\lambda)}{O_{k}(\lambda)}$ and $O_0(\lambda) := \partial_\lambda H_{\text{ad}}(\lambda)$. The nested commutators $O_{2k-1}(\lambda)$ are straightforward to compute in the spin-$1/2$ space by using the commutation relations between the Pauli matrices. However, finding the CD coefficients $\alpha_k(\lambda)$ requires an optimization step, namely the minimization of the action 
\begin{equation}
    S_\ell(\lambda) = \Tr{G_\ell^2(\lambda)},
\end{equation}
where $G_\ell(\lambda)$ is a hermitian operator defined as 
\begin{equation}
    G_\ell(\lambda) = \partial_\lambda H_{\text{ad}}(\lambda) - i \comm{H_{\text{ad}}(\lambda)}{ A_{\lambda}^{(\ell)}}.
\end{equation}
This optimization problem can be mapped to a linear system of equations~\cite{claeys2019floquet, xie2022variational}, where $\alpha_k(\lambda)$ is found from
\begin{equation}
\sum_{m=1}^{\ell} \alpha_m(\lambda) \Gamma_{m+k}(\lambda) = -\Gamma_{k}(\lambda).
\end{equation}
This relation in matrix form reads
\begin{equation}\label{eq:soegammas}
    \begin{bmatrix} \Gamma_2(\lambda) & \Gamma_3(\lambda) & \cdots & \Gamma_{\ell+1}(\lambda) \\
                    \Gamma_3(\lambda) & \Gamma_4(\lambda) & \cdots & \Gamma_{\ell+2}(\lambda) \\
                    \vdots & \vdots & \ddots & \vdots \\
                    \Gamma_{\ell+1}(\lambda) & \Gamma_{\ell+2}(\lambda) & \cdots & \Gamma_{2\ell}(\lambda)
    \end{bmatrix} \begin{bmatrix}
        \alpha_1(\lambda) \\ \alpha_2(\lambda) \\ \vdots \\ \alpha_{\ell}(\lambda) 
    \end{bmatrix} = - \begin{bmatrix}
        \Gamma_1(\lambda) \\ \Gamma_2(\lambda) \\ \vdots \\ \Gamma_{\ell}(\lambda) 
    \end{bmatrix},
\end{equation}
where $\Gamma_k(\lambda) := \norm{O_k(\lambda)}^2 = \Tr{O_k^\dagger(\lambda) O_k(\lambda)}$. Therefore, Eq. (\ref{eq:soegammas}) can be cast into $\mathcal{G}(t) \vec{\alpha}(t) = -\vec{g}(t)$. Then, the first order CD coefficient is given by 
\begin{equation}\label{eq:coefficient-form}
    \alpha_1(\lambda) = -\frac{\Gamma_1(\lambda)}{\Gamma_2(\lambda)}.
\end{equation}
For the particular Hamiltonian defined in the main text, namely $H_\text{ad} = (1-\lambda) \sum_{i} h_{i}^{x} \sigma^x_{i} + \lambda \left( \sum_{i} h_{i}^{z} \sigma^z_{i} + \sum_{i<j} J_{ij} \sigma^z_i \sigma^z_j \right)$, the term $\mathcal{O}_1(\lambda)$ takes the form
\begin{align}
\mathcal{O}_{1}(\lambda) &= \comm{H_\text{ad}(\lambda)}{\partial_\lambda H_\text{ad}(\lambda)} \\
&=\comm{H_i}{H_f}\\
&=\comm{\sum_{i} h_{i}^{x} \sigma^x_{i}+\sum_{i} h_{i}^{b} \sigma^z_{i}}{ \sum_{k} h_{k}^{z} \sigma^z_{k}+\sum_{k<l} J_{kl} \sigma^z_{k} \sigma^z_{l}} \\
&= -2i \sum_{i} h_{i}^{x} h_{i}^{z} \sigma^y_{i}-2 i \sum_{i<j} J_{i j}\left(h_{i}^{x} \sigma^y_{i} \sigma^z_{j}+h_{j}^{x} \sigma^z_{i} \sigma^y_{j}\right).
\end{align}
Therefore, 
\begin{equation}
\Gamma_{1}(\lambda) = 4 \sum_{i}\left(h_{i}^{x}\right)^{2}\left(h_{i}^{z}\right)^{2}+4 \sum_{i<j} J_{i j}^{2}\left[\left(h_{i}^{x}\right)^{2}+\left(h_{j}^{x}\right)^{2}\right].
\end{equation}
Now, we only need to calculate $\Gamma_2(\lambda)$ to find the analytical expression of $\alpha_1(\lambda)$. We start by finding the operator form of $\mathcal{O}_2(\lambda)$
\begin{equation}
\mathcal{O}_{2}(\lambda)=\left[H_\text{ad}(\lambda), \mathcal{O}_{1}(\lambda)\right]=(1-\lambda)\left[H_i(\lambda), \mathcal{O}_{1}(\lambda)\right]+\lambda\left[H_f(\lambda), \mathcal{O}_{1}(\lambda)\right].
\end{equation}
Working first with $\left[H_i(\lambda), \mathcal{O}_{1}(\lambda)\right]$
\begin{align}
\left[H_i, \mathcal{O}_{1}\right] &= \left[\sum_{i} h_{i}^{x} \sigma^x_{i}+\sum_{i} h_{i}^{b} \sigma^z_{i},-2i \sum_{j} h_{j}^{x} h_{j}^{z} \sigma^y_{j} -2 i \sum_{j<k} J_{j k}\left(h_{j}^{x} \sigma^y_{j} \sigma^z_{k}+h_{k}^{x} \sigma^z_{j} \sigma^y_{k}\right)\right]  \\
&=4 \sum_{i}\left(h_{i}^{x}\right)^{2} h_{i}^{z} \sigma^z_{i}-4 \sum_{i} h_{i}^{b} h_{i}^{x} h_{i}^{z} \sigma^x_{i}+4 \sum_{i<j}\left[\left(h_{i}^{x}\right)^{2}+\left(h_{j}^{x}\right)^{2}\right] J_{i j} \sigma^z_{j} \sigma^z_{j} \notag \\
&\quad -8 \sum_{i<j} h_{i}^{x} h_{j}^{x} J_{i j} \sigma^y_{i} \sigma^y_{j}-4 \sum_{i<j} J_{i j}\left(h_{i}^{b} h_{i}^{x} \sigma^x_{i} \sigma^z_{j}+h_{j}^{b} h_{j}^{x} \sigma^z_{i} \sigma^x_{j}\right).
\end{align}
The next step is to find an expression for $\left[H_f(\lambda), \mathcal{O}_{1}(\lambda)\right]$, as follows
\begin{align}
\left[H_f, \mathcal{O}_{1}\right]&=\left[\sum_{i} h_{i}^{z} \sigma^z_{i}+\sum_{i<j} J_{i j} \sigma^z_{i} \sigma^z_{j},-2 i \sum_{k} h_{k}^{x} h_{k}^{z} \sigma^y_{k}-2 i \sum_{k<l} J_{k l}\left(h_{k}^{x} \sigma^y_{k} \sigma^z_{l}+h_{l}^{x} \sigma^z_{k} \sigma^y_{l}\right)\right] \\
&=-4 \sum_{i}\left[\left(h_{i}^{z}\right)^{2} h_{i}^{x}+\sum_{j \neq i} J_{i j}^{2} h_{i}^{x}\right] \sigma^x_{i}-8 \sum_{i<j} J_{i j} h_{i}^{x} h_{i}^{z} \sigma^x_{i} \sigma^z_{j}-8 \sum_{i<j} J_{i j} h_{j}^{x} h_{j}^{z} \sigma^z_{i} \sigma^x_{j} \notag\\
&\quad -8 \sum_{i<j<k} J_{i j} J_{i k} h_{i}^{x} \sigma^x_{i} \sigma^z_{j} \sigma^z_{k}-8 \sum_{i<j<k} J_{j k} J_{i j} h_{j}^{x} \sigma^z_{i} \sigma^x_{j} \sigma^z_{k}-8 \sum_{i<j<k} J_{i k} J_{j k} h_{k}^{x} \sigma^z_{i} \sigma^z_{j} \sigma^x_{k}.
\end{align}
Now that we have both $\left[H_i(\lambda), \mathcal{O}_{1}(\lambda)\right]$ and $\left[H_f(\lambda), \mathcal{O}_{1}(\lambda)\right]$, we can get $\mathcal{O}_{2}(\lambda)$ as 
\begin{align}
\mathcal{O}_{2}(\lambda) &= (1-\lambda) \Bigg( 4 \sum_{i}\left(h_{i}^{x}\right)^{2} h_{i}^{z} \sigma^z_{i}-4 \sum_{i} h_{i}^{b} h_{i}^{x} h_{i}^{z} \sigma^x_{i}+4 \sum_{i<j}\left[\left(h_{i}^{x}\right)^{2}+\left(h_{j}^{x}\right)^{2}\right] J_{i j} \sigma^z_{j} \sigma^z_{j} \notag \\
&\quad-8 \sum_{i<j} h_{i}^{x} h_{j}^{x} J_{i j} \sigma^y_{i} \sigma^y_{j}-4 \sum_{i<j} J_{i j}\left(h_{i}^{b} h_{i}^{x} \sigma^x_{i} \sigma^z_{j}+h_{j}^{b} h_{j}^{x} \sigma^z_{i} \sigma^x_{j}\right) \Bigg) \notag \\
&\quad+ \lambda \Bigg( -4 \sum_{i}\left[\left(h_{i}^{z}\right)^{2} h_{i}^{x}+\sum_{j \neq i} J_{i j}^{2} h_{i}^{x}\right] \sigma^x_{i}-8 \sum_{i<j} J_{i j} h_{i}^{x} h_{i}^{z} \sigma^x_{i} \sigma^z_{j}-8 \sum_{i<j} J_{i j} h_{j}^{x} h_{j}^{z} \sigma^z_{i} \sigma^x_{j} \notag\\
&\quad -8 \sum_{i<j<k} J_{i j} J_{i k} h_{i}^{x} \sigma^x_{i} \sigma^z_{j} \sigma^z_{k}-8 \sum_{i<j<k} J_{j k} J_{i j} h_{j}^{x} \sigma^z_{i} \sigma^x_{j} \sigma^z_{k}-8 \sum_{i<j<k} J_{i k} J_{j k} h_{k}^{x} \sigma^z_{i} \sigma^z_{j} \sigma^x_{k} \Bigg).
\end{align}
Collecting common terms, 
\begin{align}
\mathcal{O}_{2}(\lambda) &= -4   \sum_{i} \left( (1-\lambda) h_{i}^{b} h_{i}^{x} h_{i}^{z} + \lambda \left[\left(h_{i}^{z}\right)^{2} h_{i}^{x}+\sum_{j \neq i} J_{i j}^{2} h_{i}^{x}\right]  \right) \sigma^x_{i} \notag \\
&\quad -4 \left( \sum_{i\neq j} (1-\lambda) J_{ij} h_i^b h_i^x + 2\lambda J_{ij} h_i^x h_i^z \right) \sigma^x_i \sigma^z_j \notag \\
&\quad + (1-\lambda) \Bigg( 4 \sum_{i}\left(h_{i}^{x}\right)^{2} h_{i}^{z} \sigma^z_{i}+4 \sum_{i<j}\left[\left(h_{i}^{x}\right)^{2}+\left(h_{j}^{x}\right)^{2}\right] J_{i j} \sigma^z_{j} \sigma^z_{j} -8 \sum_{i<j} h_{i}^{x} h_{j}^{x} J_{i j} \sigma^y_{i} \sigma^y_{j}  \Bigg) \notag \\
&\quad + \lambda \Bigg( -8 \sum_{i<j<k} J_{i j} J_{i k} h_{i}^{x} \sigma^x_{i} \sigma^z_{j} \sigma^z_{k}-8 \sum_{i<j<k} J_{j k} J_{i j} h_{j}^{x} \sigma^z_{i} \sigma^x_{j} \sigma^z_{k}-8 \sum_{i<j<k} J_{i k} J_{j k} h_{k}^{x} \sigma^z_{i} \sigma^z_{j} \sigma^x_{k} \Bigg).
\end{align}
We can distinguish between three types of terms, those with a $(1-\lambda)$ factor, those with a $\lambda$ factor and those that have contributions from both factors. We can now easily calculate $\Gamma_2(\lambda)$ since our expansion of $\mathcal{O}_2(\lambda)$ has non-repeating Pauli products.

\begin{align}
\Gamma_{2}(\lambda &)=16 (1-\lambda)^2 \sum_{i}\left(h_{i}^{x}\right)^{4}\left(h_{i}^{z}\right)^{2}+64 (1-\lambda)^{2} \sum_{i<j}\left(h_{i}^{x}\right)^{2}\left(h_{j}^{x}\right)^{2} J_{i j}^{2} \notag \\
&\quad+16 (1-\lambda)^{2} \sum_{i<j}\left[\left(h_{i}^{x}\right)^{2}+\left(h_{j}^{x}\right)^{2}\right]^{2} J_{i j}^{2} \notag\\
&\quad+ 16 \sum_{i}\left\{(1-\lambda) h_{i}^{b} h_{i}^{x} h_{i}^{z}+\lambda\left[\left(h_{i}^{z}\right)^{2} h_{i}^{x}+\sum_{j \neq i} J_{i j}^{2} h_{i}^{x}\right]\right\}^{2} \notag\\
&\quad +16 \sum_{i \neq j}\left\{J_{i j} h_{i}^{x}\left[(1-\lambda) h_{i}^{b}+2 \lambda h_{i}^{z}\right]\right\}^{2} \notag \\
&\quad +64 \lambda^2 \sum_{i<j<k} \left[J_{i j}^{2} J_{1 k}^{2}\left(h_{i}^{x}\right)^{2}+J_{j k}^{2} J_{i j}^{2}\left(h_{j}^{x}\right)^{2}+J_{i k}^{2} J_{j k}^{2}\left(h_{k}^{x}\right)^{2}\right] .
\end{align}
Hence,
\begin{equation}\label{eq:alpha}
\alpha_1(\lambda) = -\frac{\Gamma_{1}(\lambda)}{\Gamma_{2}(\lambda)} = -\frac{A}{B(1-\lambda)^{2}+C \lambda(1-\lambda)+D \lambda^{2}},
\end{equation}
where 
\begin{align}
A= & 4 \sum_{i}\left(h_{i}^{x} h_{i}^{z}\right)^{2}+4 \sum_{i \neq j}\left(h_{i}^{x}\right)^{2} J_{i j}^{2} \\
B= & 16 \sum_{i}\left(h_{i}^{x}\right)^{4}\left(h_{i}^{z}\right)^{2}+48 \sum_{i \neq j}\left(h_{i}^{x} h_{j}^{x}\right)^{2} J_{i j}^{2}+16 \sum_{i \neq j}\left(h_{i}^{x}\right)^{4} J_{i j}^{2}+16 \sum_{i \neq j} J_{i j}^{2}\left(h_{i}^{x} h_{i}^{b}\right)^{2} +16 \sum_{i}\left(h_{i}^{b} h_{i}^{x} h_{i}^{z}\right)^{2} \\
C= & 96 \sum_{i \neq j} h_{i}^{6} h_{i}^{z}\left(h_{i}^{x}\right)^{2} J_{i j}^{2}+32 \sum_{i} h_{i}^{b}\left(h_{i}^{x}\right)^{2}\left(h_{i}^{z}\right)^{3} \\
D= & 96 \sum_{i<j<k}\left(J_{i j}^{2} J_{i k}^{2}\left(h_{i}^{x}\right)^{2}+J_{j k}^{2} J_{i j}^{2}\left(h_{j}^{x}\right)^{2}+J_{i k}^{2} J_{j k}^{2}\left(h_{k}^{x}\right)^{2}\right) + 96 \sum_{i \neq j}\left(h_{i}^{x} h_{i}^{z}\right)^{2} J_{i j}^{2} \notag \\ &+16 \sum_{i}\left(h_{i}^{z}\right)^{4}\left(h_{i}^{x}\right)^{2}+16 \sum_{i \neq j} J_{i j}^{4}\left(h_{i}^{x}\right)^{2}
\end{align}


\section{Bias-field DCQO circuits}

The circuits used in BF-DCQO can be divided into two parts, the first one prepares the ground state of the initial Hamiltonian. The second one implements DCQO. In this section, we give details on both parts.

\subsection{Preparation of the initial ground state}

The smallest eigenvalue of the a single-body operator, $\left[h_i^x \sigma^x_i - h_i^b \sigma^z_i \right ]$ is given by $\lambda^{min}_i = -\sqrt{(h^b_i)^2 + (h^x_i)^2}$ and its associated eigenvector is
\begin{equation}
    \ket{\phi_i} = \phi_0 \begin{bmatrix}
    1 \\ 
    \frac{h^b_i + \lambda^{min}_i}{h^x_i}
    \end{bmatrix}.
\end{equation}
Since $h^b_i$ and $h^x_i$ are projections in $z$- and $x$-axis, the equivalent ground-state, $|\phi_i\rangle$, can be prepared by a $y$-axis rotation given by
\begin{equation}
   \theta_i = \tan^{-1}\left(\frac{h^x_i}{h^b_i - \sqrt{(h^b_i)^2 + (h^x_i)^2}}\right).
\end{equation}
Therefore, for every qubit, the ground-state can be prepared by $|\phi_i\rangle = R_y(\theta_i)|0\rangle$.

\subsection{DCQO in the impulse regime}

\begin{figure}[h]
    \centering
    \includegraphics[width=1\linewidth]{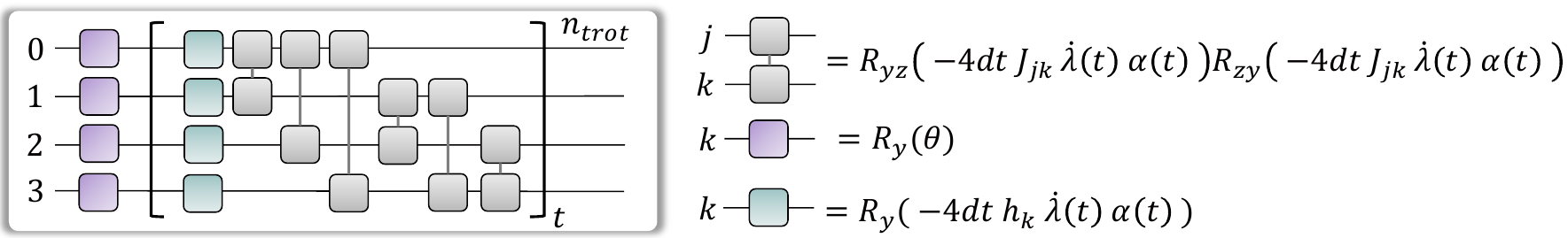}
    \caption{Structure of BF-DCQO circuits. The first layer of $R_y$ rotations prepares the initial ground state. Then, $n_\text{trot}$ Trotter steps are implemented. Within each Trotter steps, DCQO in the impulse regime is implemented. The specific gates and angles can be seen at the bottom of the figure.}
    \label{fig:sm_4qex}
\end{figure}

We restrict the expansion order of the adiabatic Gauge potential~\eqref{eq:AGP} to one. Besides, We consider the Hamiltonian evolution in the impulse regime, where the evolution time is very short. This restriction results in $|\lambda(t)| \ll |\alpha_1(t) \dot{\lambda(t)}|$, where $\alpha_1(t)$ is the first order CD coefficient. Consequently, the full Hamiltonian, i.e.\ $H_\text{ad}+\dot{\lambda}A_\lambda^{(1)}$, can be approximated using only the CD term. The BF-DCQO circuit structure for 4-qubits can be seen in Fig. (\ref{fig:sm_4qex}). 

Finally, after decomposing the unitaries of each Trotter step into one- and two-qubit gates, we omit some gates based on their angle magnitude. We choose a gate-cutoff threshold~$\theta_\text{cutoff}$, such that any gate with an angle magnitude below this threshold is discarded. Using the above techniques, we can significantly lower the circuit depth of BF-DCQO circuits while preserving algorithmic performance.

\subsection{Transpilation}
\begin{figure}[h]
    \centering
    \includegraphics[width=0.3\linewidth]{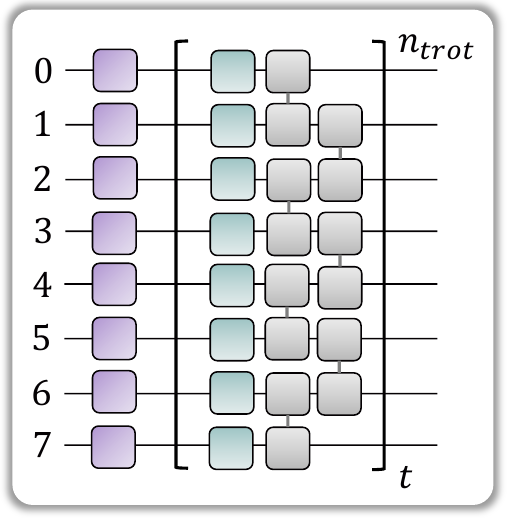}
    \caption{Parallel implementation of BF-DCQO circuits used for IBMQ's devices.}
    \label{fig:sm_parallel}
\end{figure}

Prior to running the BF-DCQO circuits on quantum hardware, we transpile those circuits into the basis gate set of the given hardware. These one- and two-qubit basis gates depend on the qubit technology as well as the hardware provider. In the case of the IonQ Forte devices~\footnote{https://ionq.com/docs/getting-started-with-native-gates}, the native gate set consists of two single-qubit unitaries
\begin{equation}
    GPi(\phi) = \begin{bmatrix}
    0 & e^{-i\phi}\\
    e^{i\phi} & 0
    \end{bmatrix}, \quad  \quad 
    GPi2(\phi) = \frac{1}{\sqrt{2}}\begin{bmatrix}
    1 & -ie^{-i\phi}\\
    -ie^{i\phi} & 1
    \end{bmatrix},
\end{equation}
and a two-qubit unitary
\begin{equation}
    ZZ(\theta) = \begin{bmatrix}
    e^{-i\frac{\theta}{2}} & 0 & 0 & 0\\
    0 & e^{i\frac{\theta}{2}} & 0 & 0 \\
    0 & 0 & e^{i\frac{\theta}{2}} & 0 \\
    0  & 0 & 0 & e^{-i\frac{\theta}{2}}
    \end{bmatrix}.
\end{equation}
These above gates are defined over continuous variables and together ensures the implementability of the relevant one- and two-body terms in our CD Hamiltonian on IonQ's trapped-ion platform, see~Table~\ref{table1}.
On the other hand, we employ the \texttt{qiskit} transpiler to construct DCQO circuits for IBMQ hardware. First we manually transpile the one- and two-body terms in our CD Hamiltonian into the following one- and two-qubit gates
\begin{equation}
    R_x(\theta) = \begin{bmatrix}
    \cos{\frac{\theta}{2}} & -i\sin{\frac{\theta}{2}}\\
    -i\sin{\frac{\theta}{2}} & \cos{\frac{\theta}{2}}
    \end{bmatrix}, \quad  \quad 
    R_y(\theta) = \begin{bmatrix}
    \cos{\frac{\theta}{2}} & -\sin{\frac{\theta}{2}}\\
    \sin{\frac{\theta}{2}} & \cos{\frac{\theta}{2}}
    \end{bmatrix}, \quad  \quad 
    R_{zz}(\theta) \equiv ZZ(\theta).
\end{equation}
Then we let the \texttt{qiskit} transpiler further optimize the circuits based on the native gates of {\it ibm\_brisbane}, namely the single qubit unitaries

\begin{equation}
    R_z(\theta) = \begin{bmatrix}
    e^{-i\frac{\theta}{2}} & 0\\
    0 & e^{i\frac{\theta}{2}}
    \end{bmatrix}, \quad  \quad 
    SX = \frac12 \begin{bmatrix}
    1+i & 1-i\\
    1-i & 1+i
    \end{bmatrix}, \quad  \quad 
    X = \begin{bmatrix}
    0 & 1\\
    1 & 0
    \end{bmatrix}
\end{equation}
and a two-qubit unitary

\begin{equation}
    ECR = \frac{1}{\sqrt{2}} \begin{bmatrix}
    0 & 1 & 0 & i\\
    1 & 0 & -i & 0 \\
    0 & i & 0 & 1 \\
    -i& 0 & 1 & 0
    \end{bmatrix}.
\end{equation}
Specifically, for IBM implementation, we apply the entangling gates in parallel, since it suits both the problem connectivity and the superconducting device capabilities. For each trotter step, there are $n-1$ entangling gates $R_{yz}(\theta) R_{zy}(\theta)$ applied between nearest neighbors. To implement all the gates we need two layers, one that applies the entangling gates between pairs of the form $(k,k+1)$, where $k$ is even; and a second layer of the form $(k,k+1)$ where $k$ is odd, see Fig. (\ref{fig:sm_parallel}) for reference. This approach allows us to reduce significantly the circuit depth by exploiting the ability of the IBMQ's devices to apply entangling gates in parallel.

\begin{table}
\caption{Transpilation of the relevant gates for a BF-DCQO circuit into IonQ Forte's native gate set and IBMQ's available gates through the qiskit SDK \cite{Qiskit}.}
\begin{tabular}{c|c|c}
\hline
Logical gate & Native decomposition for IonQ & Decomposition for IBMQ (before using the \texttt{qiskit transpiler})\\
 \hline
 \hline
 $R_y(\theta)$ & $GPi2(\pi) \times GPi(\theta/2) \times GPi2(\pi)$ & $R_y(\theta)$\\
 \hline
 $R_{zy}(\theta)$ & $(I \otimes GPi2(0))\times ZZ(\theta)\times(\theta)(I \otimes GPi2(\pi))$ & $(I\otimes R_x(\pi/2)) \times R_{zz}(\theta) \times (I\otimes R_x(-\pi/2))$\\
 \hline
 $R_{yz}(\theta)$ & $(GPi2(0)\otimes I)\times ZZ(\theta)\times(GPi2(\pi)\otimes I)$ & $(R_x(\pi/2)\otimes I) \times R_{zz}(\theta) \times (R_x(-\pi/2)\otimes I)$ \\ 
 \hline
\end{tabular}
\label{table1}
\end{table}


\subsection{Implementation}
The implementation of a randomly generated weighted maximum independent set problem on {\it IonQ-Forte 1} was carried out via Amazon Braket SDK \cite{braket}. Forte 1 is an all-to-all connected chip with 36 trapped-ion qubits. At the time of our experiments, the average one-qubit and two-qubit gate fidelities reported are 99.910\% and 99.260\%, respectively. Additionally, the qubits exhibit high coherence times exceeding one second.
Alongside these high-quality device specifications, Amazon Braket SDK provides the error mitigation technique called debiasing~\cite{maksymov2023enhancing}. This technique can be implemented simply by turning on this feature during job submission. 

We use a 127-qubit hardware \texttt{ibm\_brisbane} for running a random 100-qubit spin-glass instance, with coupling matching the heavy-hexagonal qubit connectivity of the hardware, of size 100. Due to this exact matching of the problem connectivity to the hardware, the implementation did not require any additional SWAP gates. Typical median errors for one- and two-qubit gates are 7.8e-3 and 2.44e-4, respectively, and qubit coherence was beyond $130~\mu s$.
The implementation is carried out using qiskit Runtime. We utilize the default measurement error mitigation on the sampled solution, included in optimization level 3 of the qiskit transpiler.

\section{Enhancement provided by BF-DCQO}
\subsection{Metrics}

To compare BF-DCQO with other algorithms, such as DCQO, DQA and QAOA, we employ success probability, approximation ratio and time to solution, as metrics. We define time-to-solution as 

\begin{equation}
    TTS = n_\text{iter} \cdot n_\text{shots} \cdot \frac{\log(1-0.99)}{\log(1-p_{gs})},
\end{equation}
where $n_\text{shots}$ is the number of shots used per iteration, and $n_\text{iter}$ is the number of iterations. For non-iterative algorithms $n_\text{iter}=1$. This metric reflects how much time is required to observe the ground state. Notice that the time required to make a single shot experimentally is not considered, since such times are static and hardware-dependent, i.e.\ not tunable within our algorithm. Furthermore, they are expected to improve over time.

\subsection{Simulations}

\begin{figure}[h]
    \centering
    \includegraphics[width=0.9\linewidth]{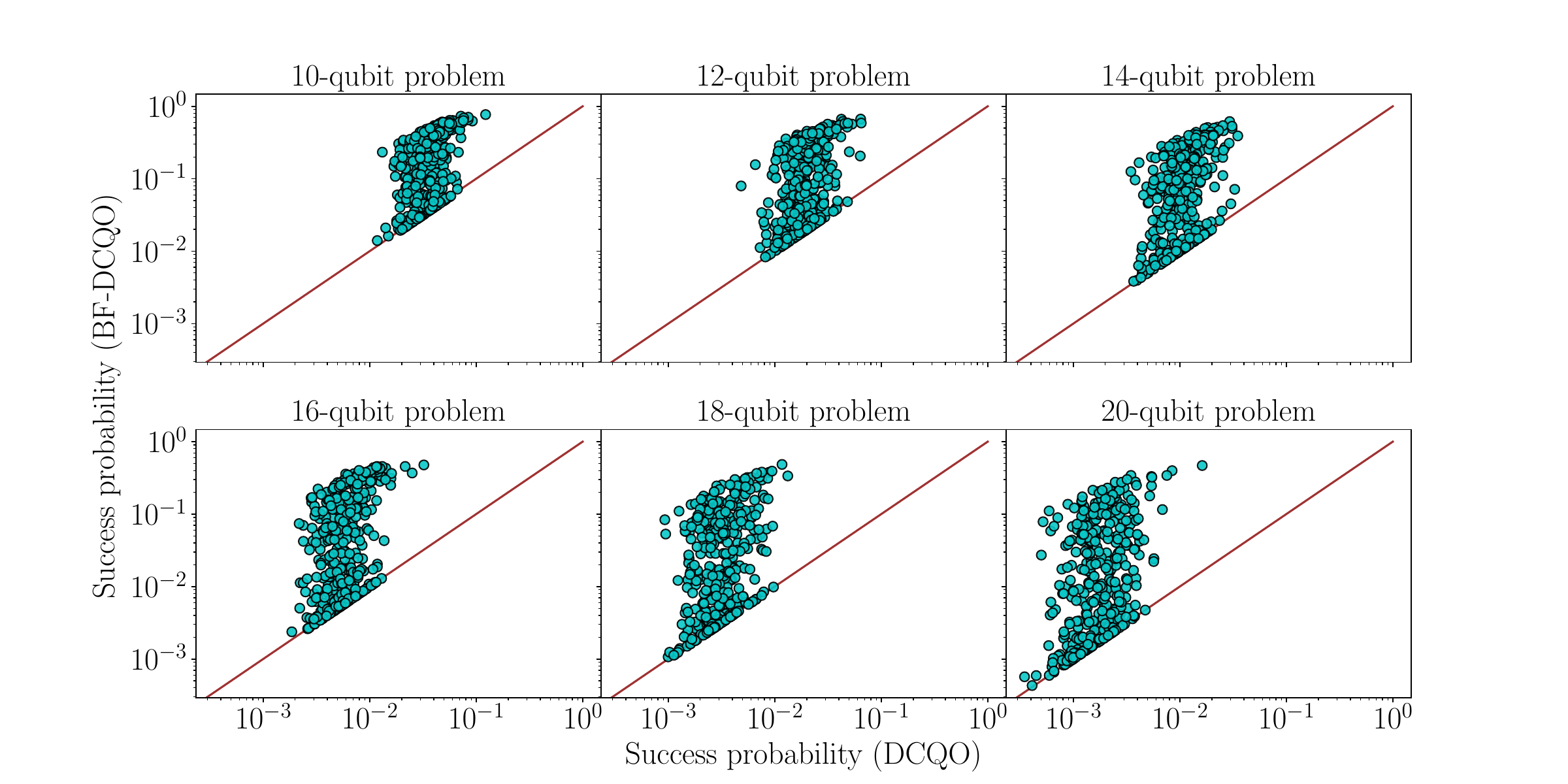}
    \caption{Success probability scatter, showing the values obtained using bias field DCQO against using only DCQO. A reference line is plotted to show how close both success probabilities are.}
    \label{fig:sm_bias_sp_scatter}
\end{figure}

\begin{figure}[h]
    \centering
    \includegraphics[width=0.9\linewidth]{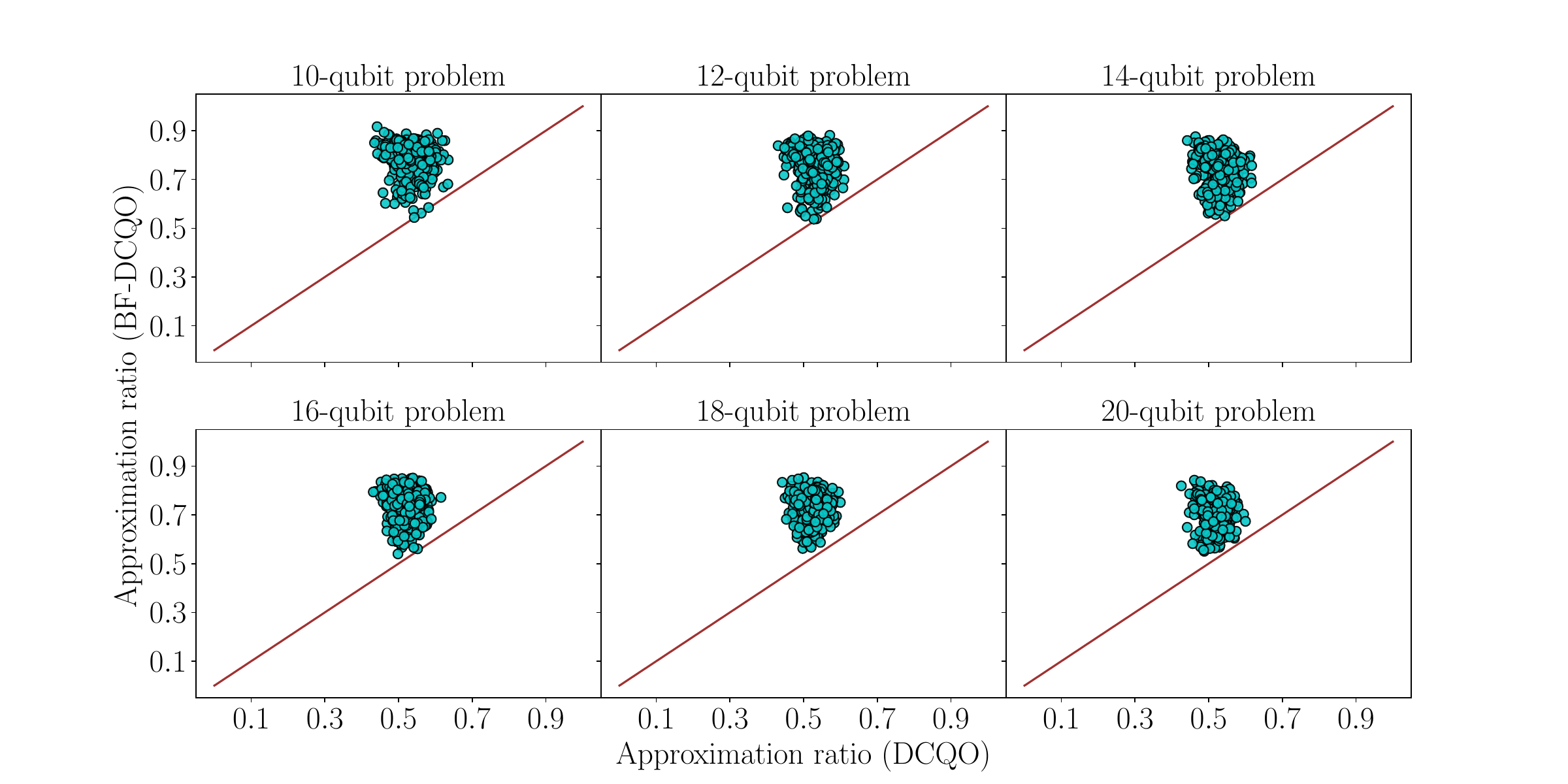}
    \caption{Approximation ratio scatter, showing the values obtained using bias field DCQO against using only DCQO. A reference line is plotted to show how close both approximation ratios are.}
    \label{fig:sm_bias_ar_scatter}
\end{figure}

\begin{figure}[h]
    \centering
    \includegraphics[width=0.9\linewidth]{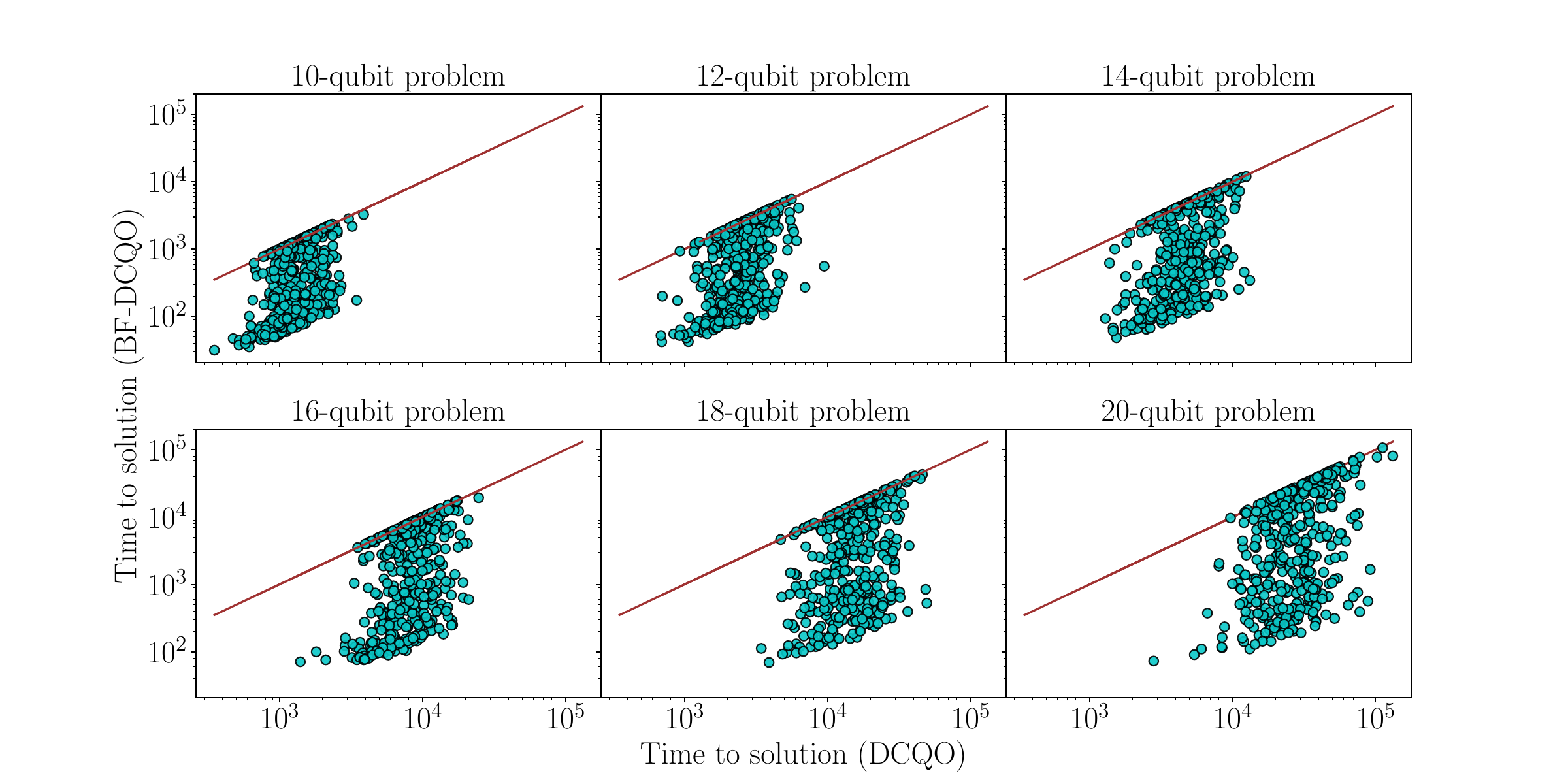}
    \caption{Time to solution scatter, showing the values obtained using bias field DCQO against using only DCQO. A reference line is plotted to show how close both times to solution are.}
    \label{fig:sm_bias_tts_scatter}
\end{figure}

\begin{figure}[h]
    \centering
    \includegraphics[width=0.5\linewidth]{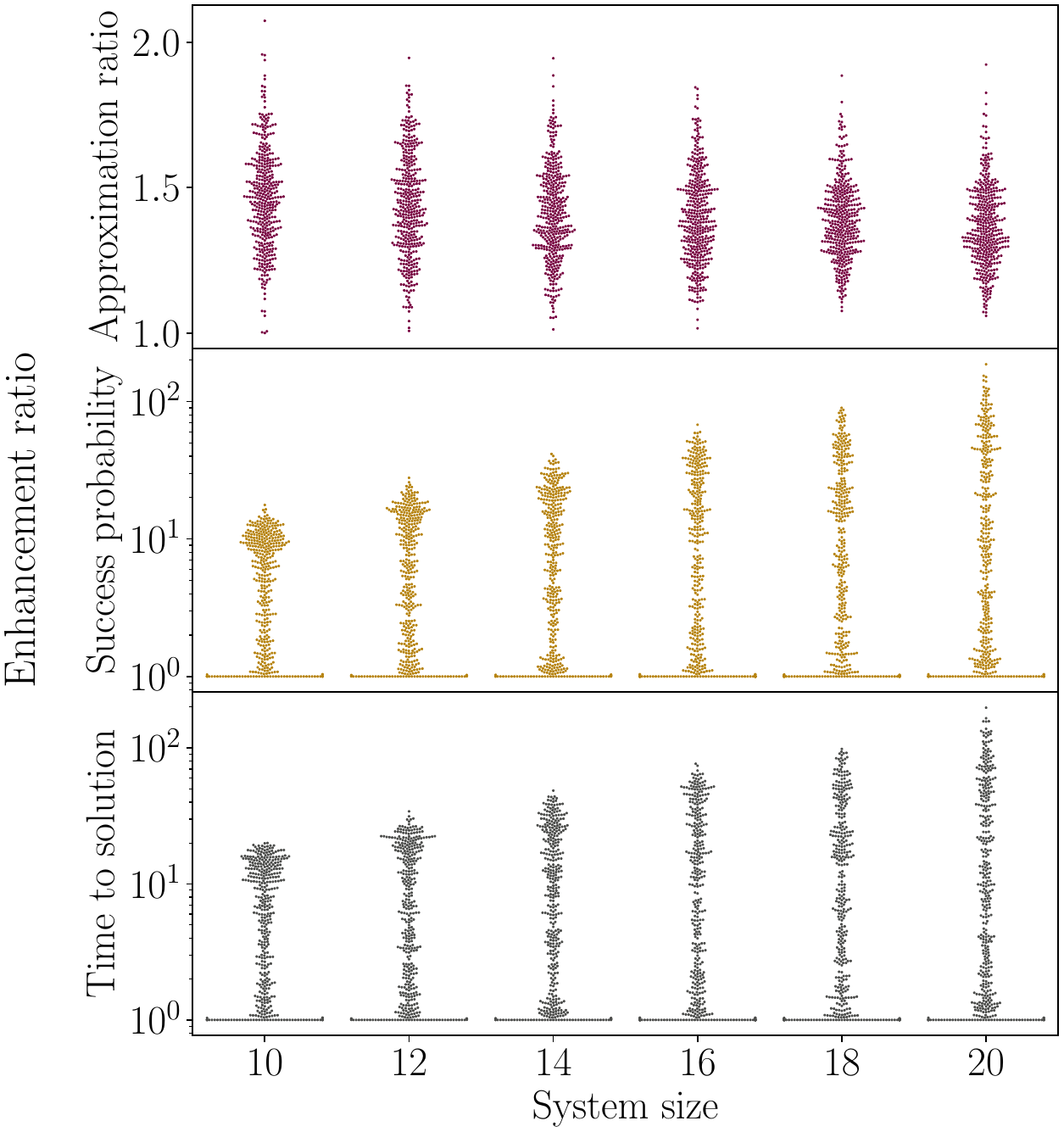}
    \caption{Enhancement ratios of bf-DQCO with respect to DCQO. For each system size and metric: success probability, approximation ratio and time-to-solution, we plot the ratios between the metric value for BF-DCQO and the value for DCQO.}
    \label{fig:sm_bias_er_scatter}
\end{figure}

We conducted a statistical study considering system sizes of 10, 12, 14, 16, 18, and 20 spins. For each system size, we simulated BF-DCQO on 400 random spin-glass instances. These instances were generated by selecting both the transverse fields $h_i$ and couplings $J_{ij}$ from a normal distribution with a zero mean and a standard deviation of one. Each instance can be retrieved using a random \texttt{Numpy} seed, ranging from 0 to 399. In our simulations, we aimed to compare BF-DCQO with DCQO. We illustrate this comparison for three metrics: success probability in Fig. (\ref{fig:sm_bias_sp_scatter}), approximation ratio in Fig. (\ref{fig:sm_bias_ar_scatter}), and time-to-solution in Fig. (\ref{fig:sm_bias_tts_scatter}). Each point in these figures corresponds to a single instance, meaning there are 400 data points for each system size. When a point lies on the reference line, it indicates that BF-DCQO performance is equal to DCQO. Additionally, in Fig. (\ref{fig:sm_bias_er_scatter}), we present the associated enhancement ratios for each metric. We observed that as the system size increases, the approximation ratio becomes less spread. This suggests that our algorithm is likely to yield better quality solutions as the system size increases, which is also evident in Fig. (\ref{fig:sm_bias_ar_scatter}), where there are no data points on the reference line for 18 and 20 spins. Moreover, in Fig. (\ref{fig:sm_bias_er_scatter}), we observe that the enhancement ratio for both success probability and time-to-solution increases as the system size increases, reflecting the polynomial enhancement observed in bf-DQCO compared to DCQO.


\begin{figure}[h]
    \centering
    \includegraphics[width=0.9\linewidth]{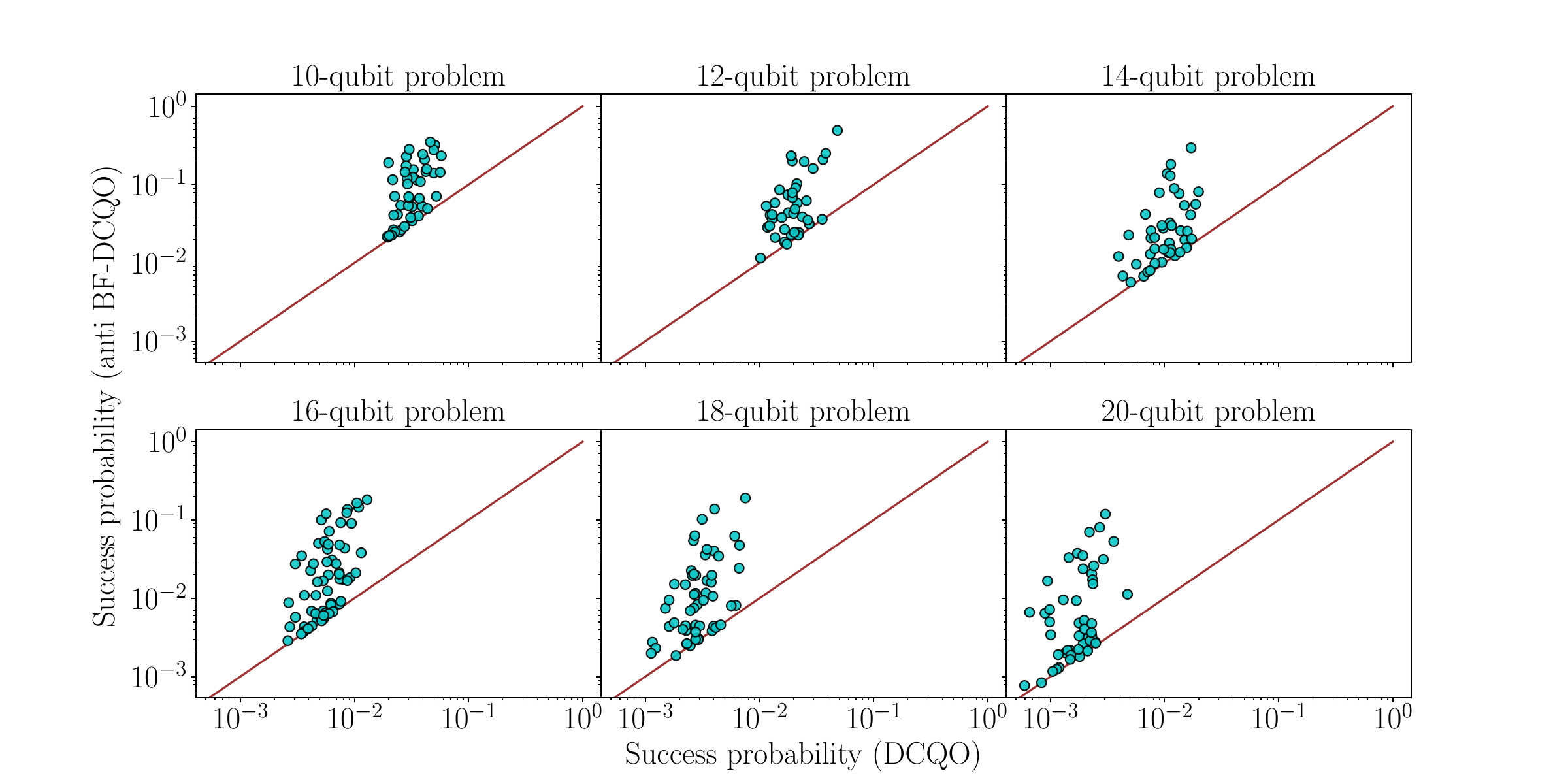}
    \caption{Success probability scatter, showing the values obtained using anti-bias field DCQO against using only DCQO, focusing on the instances in which bias field DCQO failed. A reference line is plotted to show how close both success probabilities are.}
    \label{fig:sm_antibias_sp_scatter}
\end{figure}

\begin{figure}[h]
    \centering
    \includegraphics[width=0.9\linewidth]{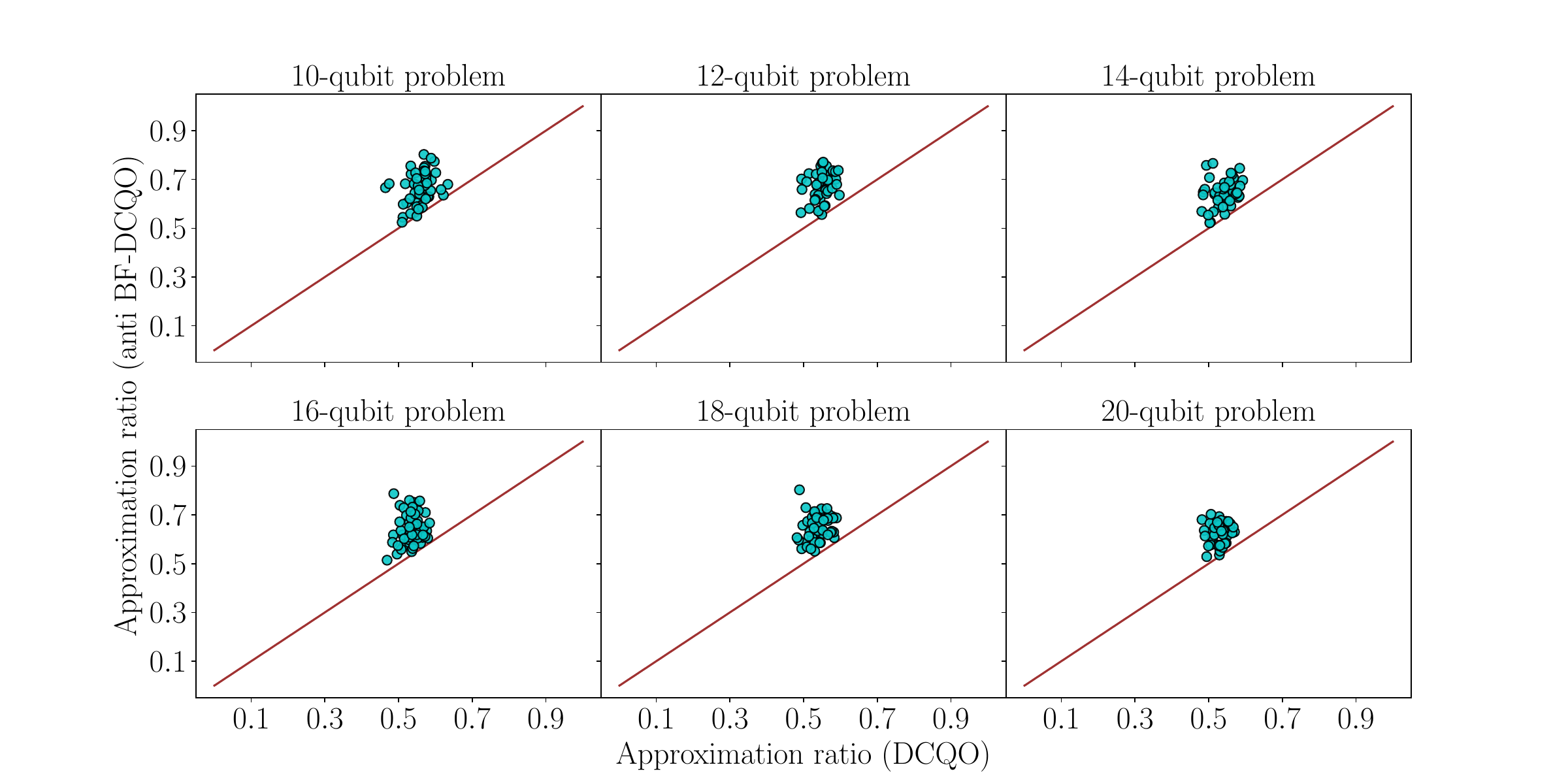}
    \caption{Approximation ratio scatter, showing the values obtained using anti-bias field DCQO against using only DCQO, focusing on the instances in which bias field DCQO failed. A reference line is plotted to show how close both approximation ratios are.}
    \label{fig:sm_antibias_ar_scatter}
\end{figure}

\begin{figure}[h]
    \centering
    \includegraphics[width=0.9\linewidth]{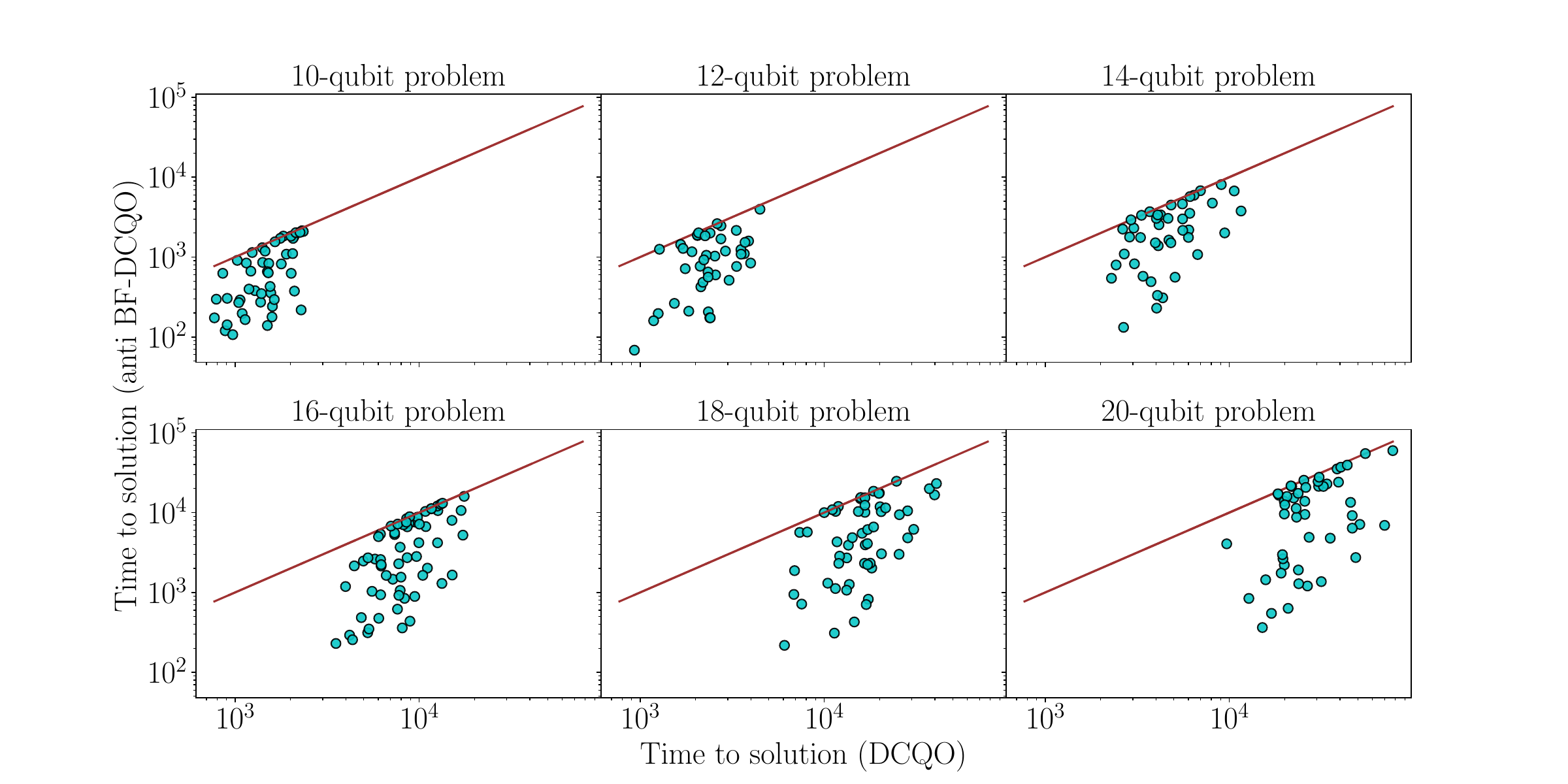}
    \caption{time to solution scatter, showing the values obtained using anti-bias field DCQO against using only DCQO, focusing on the instances in which bias field DCQO failed. A reference line is plotted to show how close both times to solution are.}
    \label{fig:sm_antibias_tts_scatter}
\end{figure}

When the initial bias fields in BF-DCQO do not effectively guide the system towards the ground state, it may deviate and localize in some of the first excited states. This deviation can lead to situations where we do not observe enhancement in success probability but still achieve a better approximation ratio. As a next step, we focused on the instances found on the reference line in Fig. (\ref{fig:sm_bias_sp_scatter}). As an alternative approach, we employed anti-bias field DCQO to counteract the direction that the system was taking with the bias fields. We present a comparison between anti-bias field DCQO and DCQO for the three metrics: success probability in Fig. (\ref{fig:sm_antibias_sp_scatter}), approximation ratio in Fig. (\ref{fig:sm_antibias_ar_scatter}), and time-to-solution in Fig. (\ref{fig:sm_antibias_tts_scatter}).


To complete our comparisons, for each system size we picked 10 instances with a good enhancement ratio for success probability and implemented standard QAOA (p=3). The QAOA algorithm is one of the state-of-the-art hybrid algorithms. It is based on a classical-quantum loop, in which a quantum computer is used to prepare a parametrized wave function, whereas the classical computer optimizes the parameters, such that the energy is minimized. We simulate this algorithm using the Tequila library \cite{tequila}.

\begin{figure}[h]
    \centering
    \includegraphics[width=0.9\linewidth]{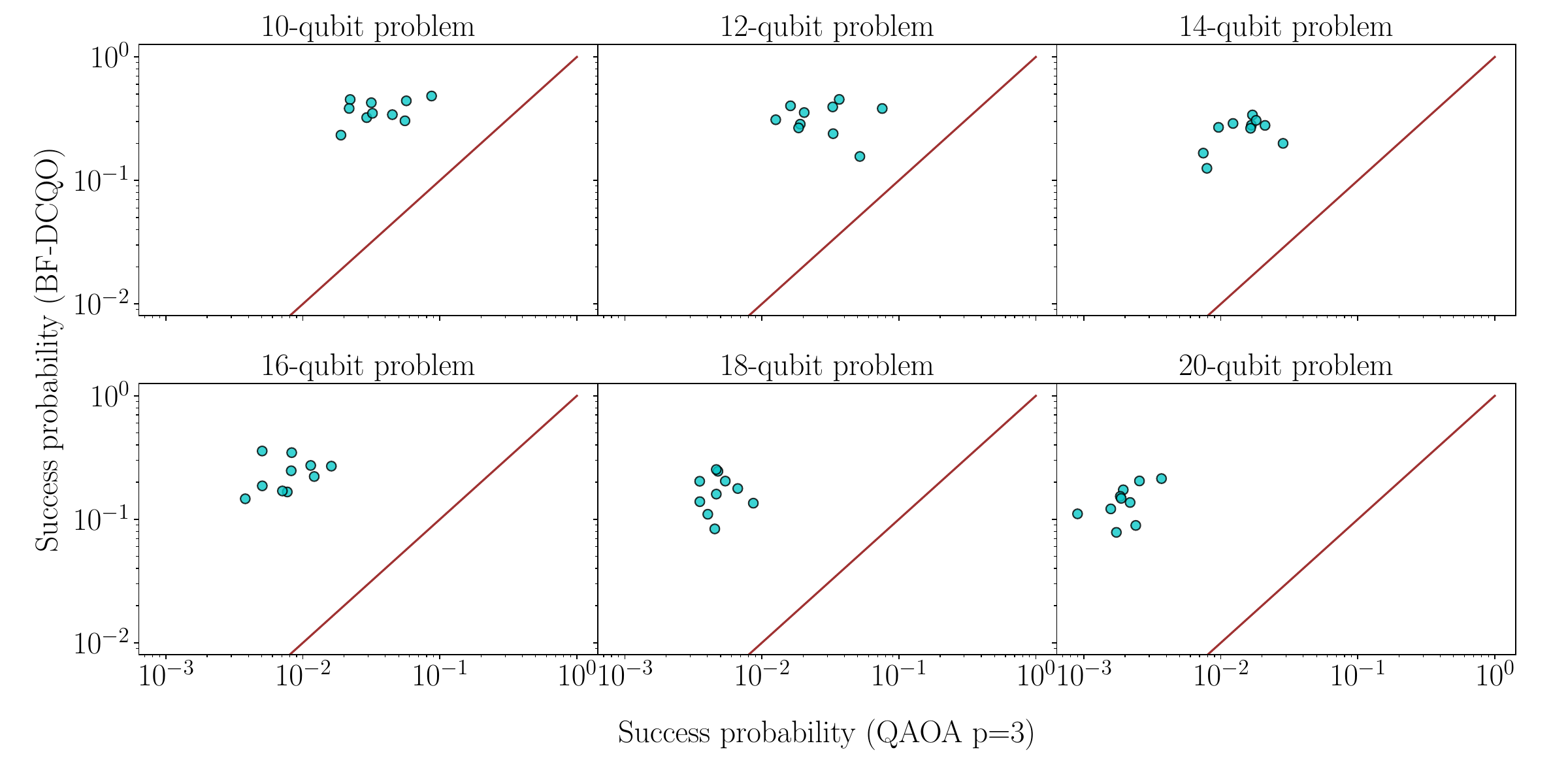}
    \caption{Success probability scatter, showing the values obtained using bias field DCQO against using QAOA (p=3). A reference line is plotted to show how close both success probabilities are.}
    \label{fig:sm_bias_qaoa_sp_scatter}
\end{figure}

\subsection{Experiments}

\begin{figure}[h]
    \centering
    \includegraphics[width=0.6\linewidth]{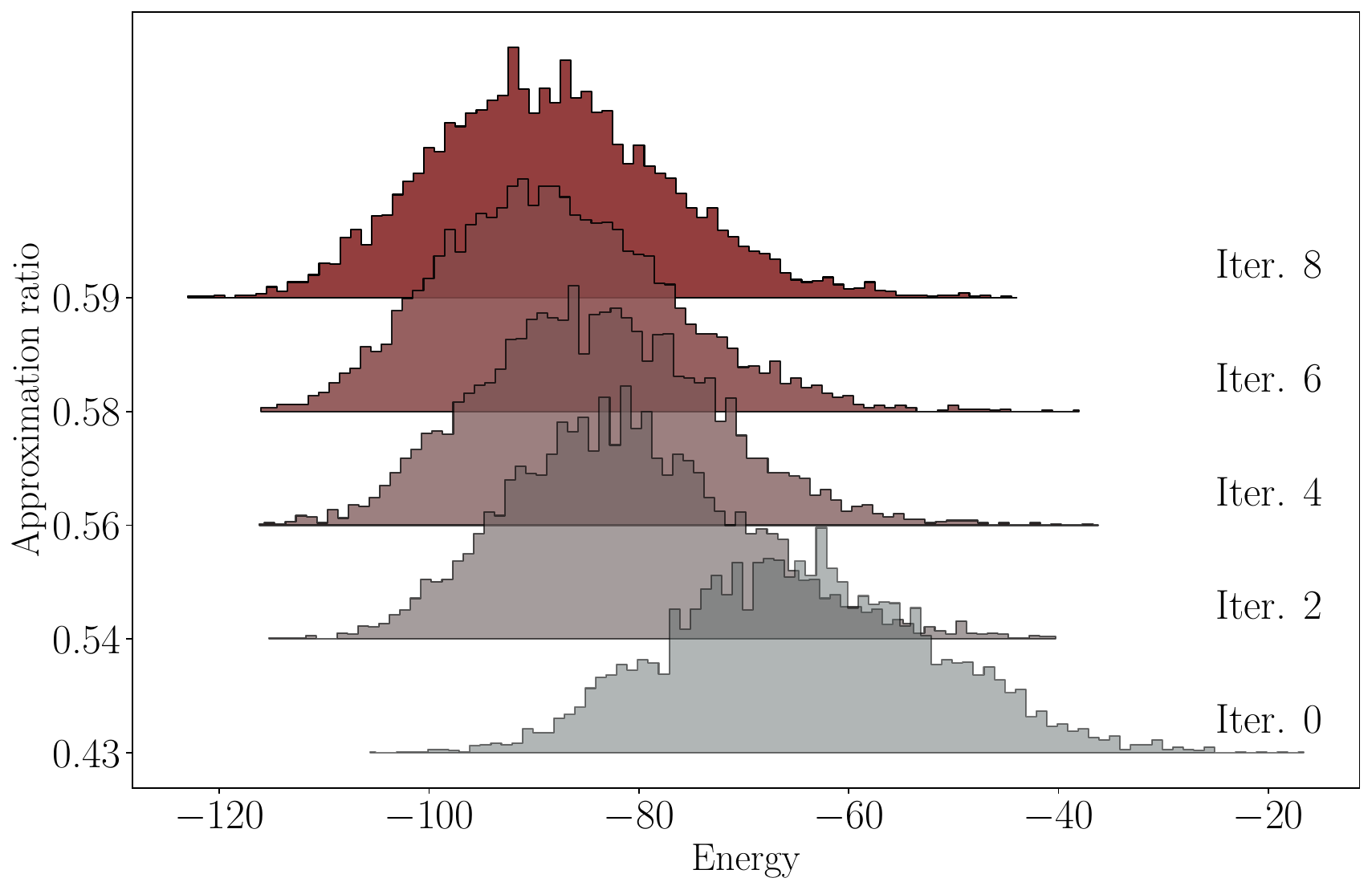}
    \caption{Purely experimental run of BF-DCQO on ibm\_brisbane, using optimization level 3 from the qiskit transpiler. Here we used $n_\text{shots}=5000$, $n_\text{steps}=2$ and $\theta_\text{cutoff}=0.05$ for each iteration. The average one-qubit gate fidelity by the time of use was $0.997$ and two-qubit gate fidelity was $0.989$. }
    \label{fig:100q-ibm-brisbane}
\end{figure}

\begin{figure}[h]
    \centering
    \includegraphics[width=0.65\linewidth]{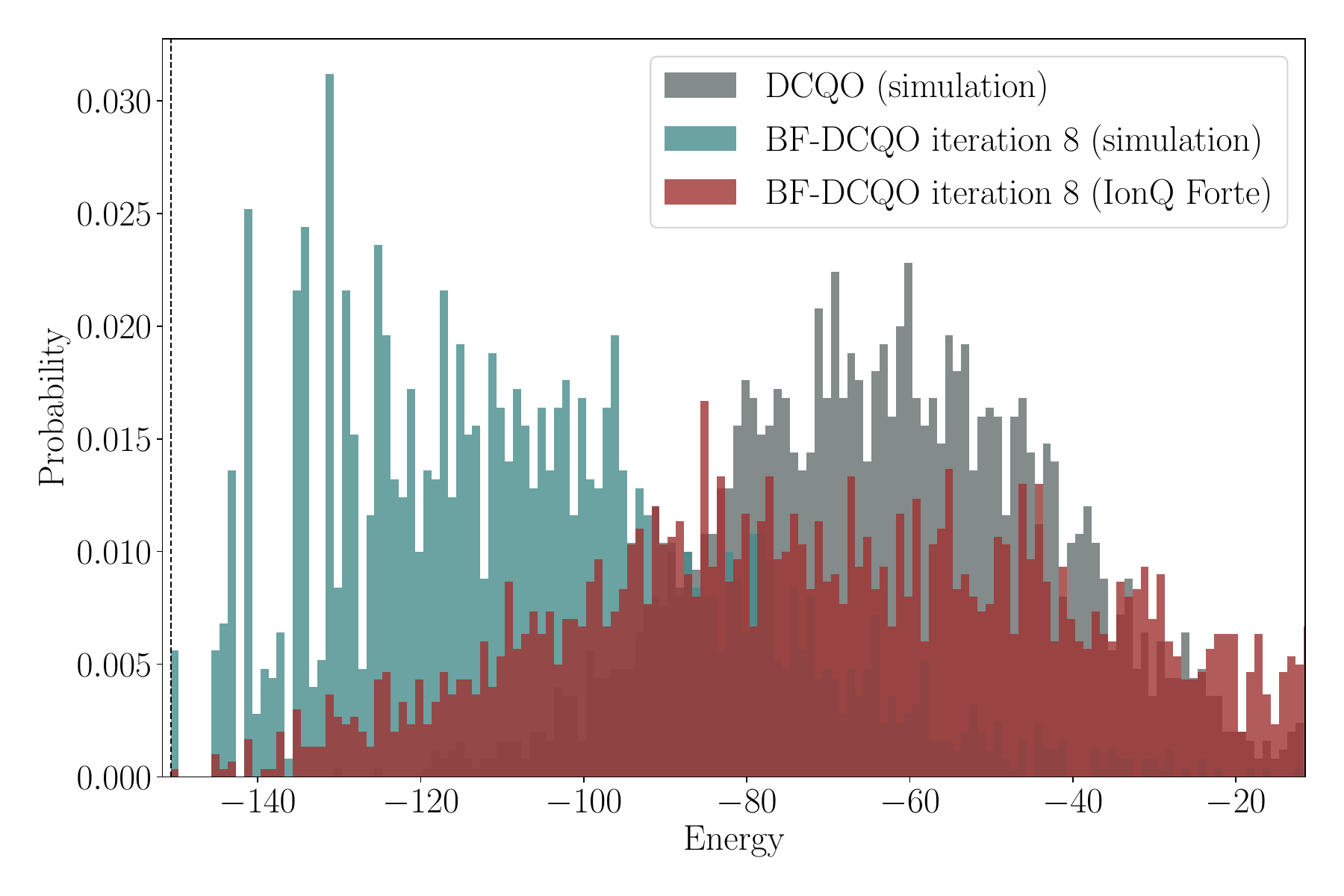}
    \caption{Energy distributions of simulated DCQO, simulated eighth iteration BF-DCQO, as well as its experimental implementation in {\it IonQ-Forte 1}. Here we used $n_\text{shots}=3000$, $n_\text{steps}=3$ and $\theta_\text{cutoff}=0.1$ for each iteration. The average one-qubit gate fidelity by the time of use was $0.9991$ and two-qubit gate fidelity was $0.9926$. We used debias error mitigation as well.}
    \label{fig:33q-ionq}
\end{figure}

We conducted additional tests on hardware to verify the performance of bf-DQCO. Firstly, we executed eight iterations of BF-DCQO on hardware for a random 100-qubit spin-glass instance, which matched the heavy-hexagonal structure from ibm\_brisbane. In Fig. (\ref{fig:100q-ibm-brisbane}), we observe that the algorithm performs effectively under noisy conditions, as the approximation ratio increases with the number of iterations. Secondly, we picked a random 33-qubit spin-glass instance with all-to-all connectivity, and implemented experimentally the eight iteration of BF-DCQO, obtaining the distributions from Fig. (\ref{fig:33q-ionq}). Despite the simulated and experimental BF-DCQO distributions being distant from each other, the experimental results yielded better success probability than DCQO, since ground state was measured once.




\end{document}